\documentclass[showpacs, oneside, twocolumn, aps,prd, nofootinbib, superscriptaddress]{revtex4-1}

\usepackage{amsmath}
\usepackage{amssymb}
\usepackage{amsfonts}
\usepackage{amssymb}
\usepackage{dcolumn}
\usepackage{bm}
\usepackage{graphicx}
\usepackage{xcolor}
\usepackage{array}

\newcommand{\be}{\begin{equation}}
\newcommand{\ee}{\end{equation}}
\newcommand{\ba}{\begin{eqnarray}}
\newcommand{\ea}{\end{eqnarray}}

\newcommand{\calR}{{\cal R}}

\begin{document}
%\title{Induced Resonant Amplification for Primordial Black Holes and Stochastic Gravitational Waves}

\rightline{YITP-20-116, IPMU20-0096}

\title{Primordial black holes and gravitational waves \\ from resonant amplification during inflation}

\author{Zihan Zhou}
\email{ustczzh@mail.ustc.edu.cn}
\affiliation{Department of Astronomy, School of Physical Sciences, University of Science and Technology of China, Hefei, Anhui 230026, China}
\affiliation{CAS Key Laboratory for Researches in Galaxies and Cosmology, University of Science and Technology of China, Hefei, Anhui 230026, China}
\affiliation{School of Astronomy and Space Science, University of Science and Technology of China, Hefei, Anhui 230026, China}

\author{Jie Jiang}
\email{jiejiang@mail.ustc.edu.cn}
\affiliation{Department of Astronomy, School of Physical Sciences, University of Science and Technology of China, Hefei, Anhui 230026, China}
\affiliation{CAS Key Laboratory for Researches in Galaxies and Cosmology, University of Science and Technology of China, Hefei, Anhui 230026, China}
\affiliation{School of Astronomy and Space Science, University of Science and Technology of China, Hefei, Anhui 230026, China}

\author{Yi-Fu Cai}
\email{yifucai@ustc.edu.cn}
\affiliation{Department of Astronomy, School of Physical Sciences, University of Science and Technology of China, Hefei, Anhui 230026, China}
\affiliation{CAS Key Laboratory for Researches in Galaxies and Cosmology, University of Science and Technology of China, Hefei, Anhui 230026, China}
\affiliation{School of Astronomy and Space Science, University of Science and Technology of China, Hefei, Anhui 230026, China}

\author{Misao Sasaki}
\email{misao.sasaki@ipmu.jp}
\affiliation{Kavli Institute for the Physics and Mathematics of the Universe (WPI),University of Tokyo, Chiba 277-8583, Japan}
\affiliation{Center for Gravitational Physics, Yukawa Institute for Theoretical Physics, Kyoto University, Kyoto 606-8502, Japan}
\affiliation{Leung Center for Cosmology and Particle Astrophysics, National Taiwan University, Taipei 10617, Taiwan}

\author{Shi Pi}
\email{shi.pi@ipmu.jp}
\affiliation{Kavli Institute for the Physics and Mathematics of the Universe (WPI),University of Tokyo, Chiba 277-8583, Japan}
\affiliation{CAS Key Laboratory of Theoretical Physics, Institute of Theoretical Physics, Chinese Academy of Sciences, Beijing 100190, China}

\begin{abstract}
We present a new realization of the resonant production of primordial black holes as well as gravitational waves 
in a two-stage inflation model consisting of a scalar field $\phi$ with an axion-monodromy-like periodic structure in the potential 
that governs the first stage and another field $\chi$ with a hilltop-like potential that dominates the second stage.
The parametric resonance seeded by the periodic structure at the first stage amplifies the perturbations of both fields inside the Hubble radius.
 While the evolution of the background trajectory experiences a turn as the oscillatory barrier height increases, 
 the amplified perturbations of $\chi$ remain as they are and contribute to the final curvature perturbation.
  It turns out that the primordial power spectrum displays a significant resonant peak on small scales, 
  which can lead to an abundant production of primordial black holes. Furthermore, gravitational waves are  
  also generated from the resonantly enhanced field perturbations during inflation, the amplitude of which 
  may be constrained by future gravitational wave interferometers.
\end{abstract}

%\pacs{98.80.Cq, 11.25.Tq, 74.20.-z, 04.50.Gh}

\maketitle

\section{Introduction}

The primordial black holes (PBHs) have become one of the crucial elements in recent cosmology \cite{zel1967hypothesis, hawking1971gravitationally, carr1974black}. 
It could serve an inspiring tool to test the unknown physics in the very early universe \cite{khlopov2010primordial, sasaki2018primordial}.
PBHs have also been considered as a candidate for dark matter (DM) \cite{carr2016primordial, carr2019primordial}. 
Moreover, with the advent of the era of gravitational wave (GW) astronomy \cite{Abbott:2016blz},
 it is suggested that observed events may be due to PBHs \cite{Bird:2016dcv}. 
Given fruitful theoretical motivations and dramatic observational developments, many efforts have been made on 
the studies of both fundamental and phenomenological aspects of PBHs including various generation mechanisms \cite{garcia1996density, Kannike:2017bxn, domcke2017pbh, kohri2018primordial, quintin2016black, fu2019primordial, Martin:2019nuw, Martin:2020fgl, cai2020primordial}. 

As one of generation mechanisms, it was suggested that the power spectrum of the primordial curvature perturbation might be
 resonantly enhanced via the sound speed resonance (SSR) 
 mechanism \cite{Cai:2018tuh} 
or through non-conventional couplings \cite{fu2019primordial}. 
However, since the comoving curvature perturbation on super-Hubble scales is conserved 
if there is only a single adiabatic mode \cite{lyth2005general, langlois2005conserved},
there is no growth of the curvature perturbation on super-Hubble scales in such a case. This is a fairly
strong requirement on the model construction.
Thus, if one wants to realize a controllable instability for the phenomenological purpose of inflationary cosmology,
 it is simplest to add a second field that plays the role of an entropy mode. Then, one may be curious about whether and under what circumstances such an instability may take place during inflation and
  the amplified entropy perturbation could eventually be converted into the curvature perturbation.

To address the above issue, we in this paper put forward a novel realization of resonantly produced PBHs through a two-field model of inflation 
in which there are two stages: 
The first stage is dominated by a field $\phi$ having a small oscillatory feature in the potential, like in monodromy inflation,
and the second stage is dominated by another field $\chi$ with a hill-top like potential.
The earlier times of the first stage account for the observed cosmic microwave background (CMB) experiments 
with no effect of the oscillatory feature.
The oscillatory feature is assumed to become gradually more significant at later times of the first stage until it stops the motion of $\phi$
at one of the minima of the oscillatory potential. During the first stage $\chi$ remains essentially massless and hence has no dynamics, 
while it starts rolling down the potential hill after $\phi$ stops evolving. 
At the last few $e$-folds of the first stage, the oscillations in $\phi$ induce a resonant amplification of the field fluctuation $\delta\phi$
inside the Hubble radius, which subsequently leads to an amplification of the field fluctuation $\delta\chi$.
In the field space, the background evolution is along the $\phi$-direction at first and then turns its trajectory to the $\chi$-direction.
This implies that the entropy perturbation at the first stage, $\delta\chi$, is converted into the curvature perturbation at the second stage.
Thus one expects a resonant peak to appear in the primordial curvature perturbation power spectrum on small scales.
Then it will naturally lead to the production of abundant PBHs. 

In addition to the PBH production, it is important to note that a substantial amount of GWs can be
produced by the scalar-scalar-tensor coupling \cite{baumann2007gravitational}.
 Accordingly, the induced GWs can be used to search for PBHs \cite{Saito:2009jt, bugaev2011constraints, Inomata:2016rbd, garcia2017gravitational, Carr:2017jsz} 
 as well as to obtain observational constraints on the early universe models \cite{bullock1997non, ivanov1998nonlinear}.
  Commonly the induced GWs are considered to be those generated during the radiation-dominated epoch
  when PBHs are formed \cite{Saito:2009jt, bugaev2011constraints, Inomata:2016rbd, garcia2017gravitational}. 
  However, it was recently noted that, the contribution of those from an inflationary epoch could be significant in certain circumstances, such as with speed of sound lower than unity \cite{biagetti2013enhancing}, with an oscillating speed of sound for scalar \cite{Cai:2019jah} and tensor modes \cite{Cai:2020ovp}.
In fact, in the model considered in the present study, we observe that the contribution of GWs induced during inflation can dominate the cosmological GW background.  The resulting GW energy spectrum is expected to be detected by future GW experiments. 

The article is organized as follows. In Sec.~\ref{IRA}, we put forward a specific, simple model of two-field inflation, and 
 present detailed analyses of the background evolution and the resonance effect on the field fluctuations induced by the oscillatory potential.
 In Sec.~\ref{PBH}, we study the PBH formation and derive the corresponding mass spectrum in our model. 
 In Sec.~\ref{GWs}, we study the GWs induced by primordial scalar perturbations during both inflationary epoch and radiation-dominated epoch.
Then we derive the GW energy spectrum that may be of observational interest in various GW surveys. 
Finally, we conclude with discussion in Sec.~\ref{Con}. Some detailed but lengthy calculations for our two-field model are presented in 
Appendices A $\sim$ D.

Throughout the article, we work in the natural units $c=\hbar=1$ and adopt the metric signature $(-,+,+,+)$. 
The reduced Planck mass is defined as $M_p=1/\sqrt{8\pi G}$. 
The dot denotes the cosmic proper time derivative ($\dot{}=d/dt$) and the prime denotes the conformal time derivative (${}'=d/d\tau$)
where $d\tau=dt/a(t)$.

\section{resonant amplification of cosmological perturbations}
\label{IRA}

We begin this section with a brief discussion on a scenario of the very early universe that involves a potential with oscillatory modulations. 
This scenario may arise from new physics beyond the standard model such as the  combination of axion monodromy inflation \cite{Silverstein:2008sg, McAllister:2008hb} and relaxion mechanism \cite{graham2015cosmological, espinosa2015cosmological}. Many extensions of the relaxion model, like the relaxion stopping mechanism \cite{nayara2020relaxion,ibe2019fast}, non-inflationary relaxion \cite{nayara2020relaxion,fonseca2018higgs} and relaxion as inflaton \cite{tangarife2018dynamics}, have been widely researched.
Here we extend such a model to a two-field model to dramatically enhance the effect of the oscillations.

The inflationary era can be divided into two stages.
The first stage is $\phi$-dominated with the effect of the oscillations become increasingly more significant as time goes on.
At early times of the first stage where the oscillatory feature is negligible, we obtain the conventional nearly scale-invariant power 
spectrum of the primordial curvature perturbation in agreement with the CMB experiments \cite{Akrami:2018odb}.

At the last couple of $e$-folds of the first stage, the periodic oscillations in the potential lead to a Mathieu-like equation 
for the field fluctuation $\delta\phi$ inside the Hubble radius and parametrically amplify it.
 It turns out that thus amplified $\delta\phi$ can induce a large enhancement of $\delta\chi$ on sub-Hubble scales. 

After $\phi$ is stopped at one of the minima of the oscillatory potential, the second stage of inflation
commences where the evolution is governed by $\chi$. 
Consequently, $\delta\chi$ generated during the first stage is converted to the adiabatic perturbation after Hubble crossing
and gives rise to the curvature perturbation. This implies that the power spectrum on small scales is dominantly determined by
the field fluctuation $\delta\chi$.

As a side remark, one may also introduce the $\chi^2\Psi^2$ term where $\Psi$ could represent for some of the SM fields,
so that the model can successfully experience reheating of the universe.

\subsection{The model}
\label{PhysicalModel}

\begin{figure}[h!]
\centering
\includegraphics[width=0.45\textwidth]{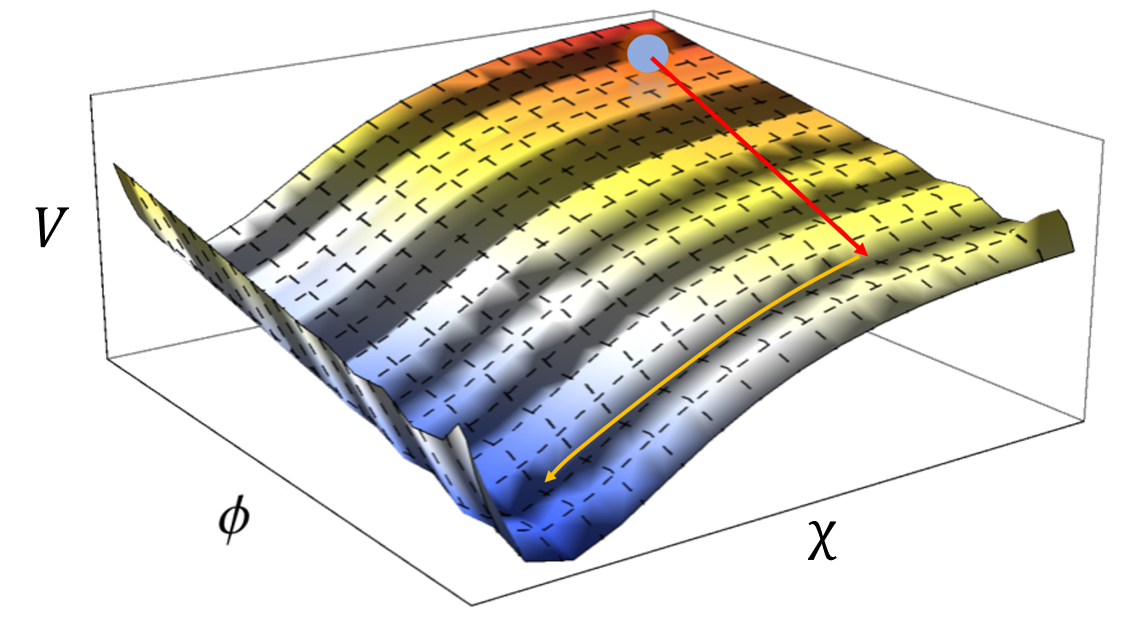}
\caption{A schematic diagram of the potential $V(\phi,\chi)$. 
	The bright and dark stripes respectively represent the local maxima and minima in the field space. 
	The red and yellow arrowed lines describe the background trajectory of the inflationary universe.}
\label{fig:Potential}
\end{figure}

The potential $V(\phi,\chi)$ of our model is sketched in Fig.~\ref{fig:Potential}.
For simplicity and concreteness, we assume $\phi$ to have a linear potential with small periodic oscillations
 and $\chi$ to have a pure linear potential. 
 To be specific, we take the Lagrangian of the form,
\begin{equation}\label{Lagrangian}
\mathcal{L} = -\frac{1}{2}(\partial_{\mu}\phi)^2 -\frac{1}{2}(\partial_{\mu}\chi)^2 -V(\phi,\chi) ~,
\end{equation}
with the two-field potential given by
\begin{equation}\label{Potentialeq}
V(\phi,\chi) = g \Lambda_{0}^3\phi +\Lambda^4(\phi){\rm cos}\big(\frac{\phi}{f_a}\big) +\xi\Lambda_{0}^3\chi +V_0 ~.
\end{equation}
In the above, the dimensional coefficient $\Lambda_0$ determines the characteristic energy scale for the background evolution. 
The dimensionless coefficients $g$ and $\xi$ are coupling constant which generate the slope. The mass scale $f_a$ determines the period of the oscillatory feature, and $\Lambda(\phi)$ describes 
a field-dependent amplitude of barriers, which we set as
\begin{equation}\label{Barrier}
\Lambda(\phi) = \Lambda_0 (1 + \alpha \frac{\phi}{M_p}) ~.
\end{equation}
In order to characterize the modulation of the potential from the periodic barriers,
 we define the monotonicity parameter $b_*$,
\begin{equation}
b_{*}(\phi)=\frac{\Lambda^4(\phi)}{|g|\Lambda_0^3f_a} ~.
\end{equation}
This parameter is also useful for analyzing the background vacuum stability in the $\phi$ direction, 
which is discussed in Appendix~\ref{Stable}.

\begin{figure}[h!]
\centering
\includegraphics[width=0.45\textwidth]{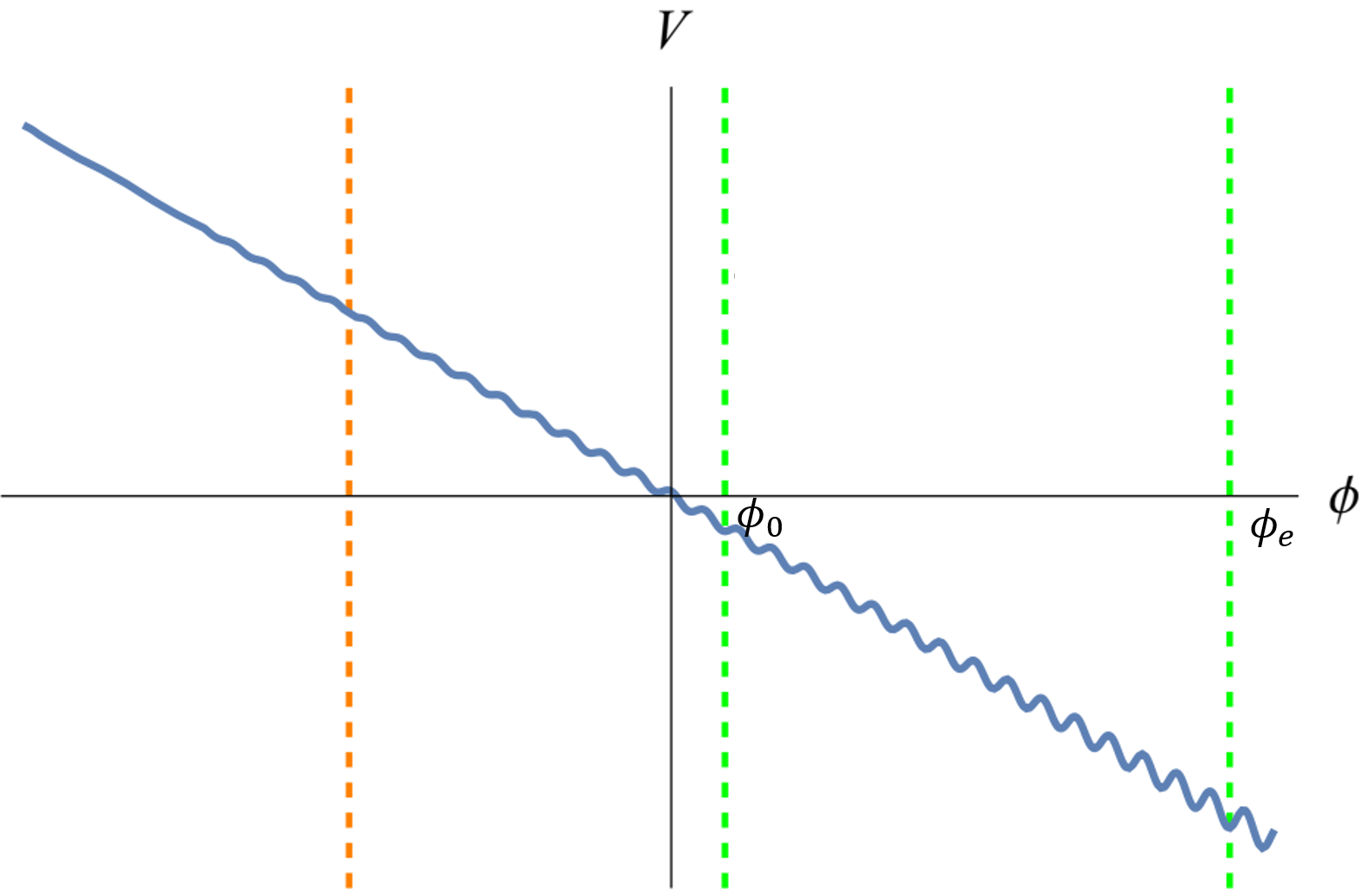}
\caption{The large panel depicts the $\phi$-dependent part of the potential. 
	The evolution proceeds in the direction of $\phi$.	
	The orange vertical line represents an epoch when the $\Lambda$ barrier terms become non-negligible,
	and $\phi_0$ represents the time at which the resonance begins. The rolling of $\phi$ stops at $\phi_e$. 
}
\label{fig:Potentialphi}
\end{figure}

The detailed feature of the $\phi$-dependent part of the potential is shown in Fig.~\ref{fig:Potentialphi}.
 On the left side of the potential, the barriers are small enough so that the potential can be viewed as a linear inflationary potential 
 with only the higher order slow-roll conditions being violated. This guarantees the scale-invariant spectrum on large scales. 
After the field rolls down and pass through the point $\phi_0$, the barrier $\Lambda(\phi)$ becomes relevant for 
certain comoving wave numbers $k$ to fall into the resonant band. 
Then, after some excursion of the field, the barrier is so large that $\phi$ stops rolling and stabilized at $\phi=\phi_e$.
After that $\chi$ drives inflation and reheating. 
Here, we also set the field excursion $\Delta\phi=\phi_e-\phi_0$, to be small enough so that $|g|\Lambda_0^3\Delta\phi\ll V_0$, 
and thus making $V_0$ play the role of cosmological constant. This considerably simplifies the analysis of the power spectrum.

In addition to the above, we set the following theoretical constraints on the model parameters:
	\begin{align}
	\bullet~&\mbox{The evolution of $\phi$ dominates first:}\nonumber\\
	&\quad |g|\gg|\xi|\,.
	\label{phidom}\\
	\bullet~&\mbox{A sufficiently flat potential at the early stage:}\nonumber\\
	&\quad b_*(\phi\ll\phi_0)\ll1\,;
	\label{flatpot}\\
	\bullet~&\mbox{A successful mechanism to stop the rolling of $\phi$:}\nonumber\\
	&\quad b_*(\phi_e)\gtrsim1\,;
	\label{stop}\\
	\bullet~&\mbox{Classical slow rolling beats quantum fluctuations:}\nonumber\\
	&\quad \frac{\dot\phi}{H}>\frac{H}{2\pi}\,;
	\label{classical}\\
	\bullet~&\mbox{The effective mass for $\delta\phi$ on flat slicing is dominated}\nonumber\\
	&\mbox{by $V''(\phi)$ when the parametric resonance happens:}\nonumber\\
	&\quad V_0\gg2|g|\Lambda_0^3f_a\,;
	\label{effmass}
\end{align}

\subsection{Background evolution}

Now we present a semi-analytical study of the background evolution.
 The Friedmann equations for the background are
\begin{equation}\label{Friedmann}
H^2=\frac{\rho}{3M_{p}^2} ~,~~ 
\frac{\ddot a}{a}=-\frac{1}{6M_p^2}(\rho+3p) ~,
\end{equation}
where the background energy density and pressure are given by
\begin{equation}\label{denpre}
\rho=\frac{1}{2}\dot{\chi}^2+\frac{1}{2}\dot{\phi}+V(\phi,\chi) ~,~~ p=\frac{1}{2}\dot{\chi}^2+\frac{1}{2}\dot{\phi}-V(\phi,\chi) ~.
\end{equation}
The Klein-Golden equations for $\phi$ and $\chi$ are 
\begin{equation}\label{BGKG}
\ddot{\phi}+3H\dot{\phi}+\frac{\partial V}{\partial \phi}=0 ~,~~ \ddot{\chi}+3H\dot{\chi}+\frac{\partial V}{\partial \chi}=0 ~.
\end{equation}

For convenience, we introduce a set of slow-roll parameters as follows:
\begin{align}
\begin{split}
\epsilon_{\phi\phi}=\frac{\dot\phi^2}{2H^2M_p^2} &, \quad \epsilon_{\chi\chi}=\frac{\dot\chi^2}{2H^2M_p^2}\,,
\\
\eta_{\phi\phi}=\frac{\dot\epsilon_{\phi\phi}}{H\epsilon}&,\quad
\eta_{\chi\chi}=\frac{\dot\epsilon_{\chi\chi}}{H\epsilon}\,,
\end{split}
\end{align} 
where it is easy to see that $\epsilon = -\dot H/H^2 = \epsilon_{\phi\phi} +\epsilon_{\chi\chi}$. 

Then, assuming all these slow-roll parameters are small, we obtain expressions for $\ddot\phi$, $\dot\phi$ and $\dot\chi$ in terms of $\phi$,
under the slow-roll approximation,
\begin{align}\label{BackEoM}
\ddot{\phi} &=3Hb_*(\phi)\dot\phi_0\sin(\frac{\phi}{f_a})\,, \nonumber\\
\dot{\phi} &= \dot\phi_0-3Hf_a b_*(\phi)\cos(\frac{\phi}{f_a})\,,\nonumber\\
\dot{\chi} & =\dot\chi_0\equiv -\frac{\xi \Lambda_0^3}{3 H}\,.
\end{align}
where
\begin{equation}
	\label{dotphi0}
	\dot\phi_0\equiv-\frac{g\Lambda_0^3}{3H} \,.
\end{equation}
During the initial $\phi$-dominant stage, the monotonicity parameter $b_*$ is small enough so that  
we have $\dot\phi\simeq\dot\phi_0$ and $\dot\chi\ll\dot{\phi}$. 
In particular, this implies that the evolution of $\chi$ can be ignored at this stage.

In passing, we mention that it may be of interest to compute a higher order slow-roll parameter $\dot{\eta_{\phi\phi}}$,
\begin{equation}
\frac{\dot\eta_{\phi\phi}}{H} \simeq 6 b_*{\rm cos}(\frac{\phi}{f_a}) \frac{\sqrt{2 \epsilon_{\phi \phi}}}{f_a/M_p}\,.
\end{equation}
The value of this parameter might be crucial for evaluating the primordial non-Gaussianity 
as the cubic Lagrangian will contain terms proportional to it.

As $\phi$ grows at later times of the first stage, the second oscillatory term in $\dot\phi$ in Eqs.~(\ref{BackEoM})
gradually becomes comparable with the first term, and eventually gives a barrier high enough to stop the
evolution at $\phi=\phi_e$. Then the second stage of inflation begins with $\chi$ governing the evolution.

\subsection{Field fluctuations and induced amplification}\label{PertAmplification}

To analyze the resonant amplification of the curvature perturbation, we need to properly calculate the field fluctuations.
We give a basic review of necessary formulas in Appendix~\ref{MInflation}.

In terms of the field fluctuations on flat slicing $\delta\phi^{I}$, the perturbation equations are written as
\begin{align}\label{PertEoM}
\begin{split}
\ddot{\delta\chi}_k+3H\dot{\delta\chi}_k+\frac{k^2}{a^2}\delta\chi_k+m_{\chi\chi}^2\delta\chi+m_{\chi\phi}^2\delta\phi=0\,, \\
\ddot{\delta\phi}_k+3H\dot{\delta\phi}_k+\frac{k^2}{a^2}\delta\phi_k+m_{\phi\phi}^2\delta\phi_k+m_{\chi\phi}^2\delta\chi_k=0\,, \\
m_{\chi\chi}^2=\frac{\partial^2V}{\partial\chi^2}-\frac{1}{M_{p}^2}\bigg(3\dot{\chi}^2+\frac{2\dot\chi\ddot\chi}{H}-\frac{\dot H\dot\chi^2}{H^2}\bigg)\,, \\
m_{\phi\phi}^2=\frac{\partial^2V}{\partial\phi^2}-\frac{1}{M_{p}^2}\bigg(3\dot{\phi}^2+\frac{2\dot\phi\ddot\phi}{H}-\frac{\dot H\dot\phi^2}{H^2}\bigg)\,, \\
m_{\chi\phi}^2=\frac{\partial ^2V}{\partial\chi\partial\phi}-\frac{1}{M_{p}^2}\bigg(3\dot\chi\dot\phi+\frac{\dot\chi\ddot\phi+\dot\phi\ddot\chi}{H}-\frac{\dot H\dot\chi\dot\phi}{H^2}\bigg)\,, 
\end{split}
\end{align}
Noting the conditions (\ref{phidom}), (\ref{flatpot}), and (\ref{effmass}) imposed on our model,
the relative magnitude among the effective mass squares is found as $m_{\phi\phi}^2\gg m_{\chi\phi}^2\gg m_{\chi\chi}^2$. 
Then the above equations are simplified to be
\begin{align}
&\ddot{\delta\chi}_k+3H\dot{\delta\chi}_k+\frac{k^2}{a^2}\delta\chi_k\simeq\frac{\dot\chi\ddot\phi}{M_{p}^2H}\delta\phi_k\,,
\label{EoMQchi}\\
&\ddot{\delta\phi}_k+3H\dot{\delta\phi}_k+\bigg(\frac{k^2}{a^2}-\frac{\Lambda^4(\phi)}{f_a^2}{\rm cos}(\frac{\phi}{f_a})\bigg)\delta\phi_k=0\,.
\label{EoMQphi}
\end{align}

First we consider the equation for $\delta\phi_k$. 
Introducing a new variable ${\delta\Phi}_k=a^{3/2}(t){\delta\phi}_k$, Eq.~(\ref{EoMDPhi}) becomes
\begin{equation}\label{EoMDPhi}
\ddot{\delta\Phi}_k+\omega_k^2(t)\delta\Phi_k=0\,,
\end{equation} 
where
\begin{equation}\label{EoMDPhiOmega}
\omega_k^2(t)=\frac{k^2}{a^2(t)}-\frac{\Lambda^4(\phi)}{f_a^2}{\rm cos}(\frac{\phi}{f_a})-\frac{9}{4}H^2-\frac{3}{2}\dot H \,.
\end{equation}
On sub-Hubble horizon scales, and for an interval of time smaller than the Hubble expansion time, 
the scale factor can be regarded as adiabatically slowly varying, with $H^2\sim const.$ and negligible $\dot H$.
Thus the above equation reduces to the Mathieu equation in the expanding universe background. 
A detailed analysis of this equation is given in Appendix~\ref{ResonanceInExpand}. The resonant amplification gives
\begin{equation}\label{ExpQphi}
	|\delta\phi_k|\propto\exp\Big[\lambda_k Ht\Big]\,,
\end{equation}
where the growth rate $\lambda_k$ is given by
\begin{equation}
\lambda_k=\mu_{k}\frac{|g|\Lambda_0^3}{6H^2f_a}-\frac{3}{2}\,,
\label{growthrate}
\end{equation}
with $\mu_k$ being the Floquet number for the $k$ mode.

We now turn to the equation for $\delta\chi_k$, Eq.~(\ref{EoMQchi}), and focus on the most enhanced mode $k_*$ 
which crosses the horizon just at $\phi=\phi_e$. As long as the source term proportional to $\delta\phi_k$ is negligible, 
that is, when $|\frac{\dot\chi\ddot\phi}{M_{p}^2H}\delta\phi_{k_*}|\ll|\frac{k^2}{a^2}\delta\chi_{k_*}|$,
which holds at $\phi\sim\phi_0$, $\delta\chi_{k_*}$ decays as 
\begin{equation}\label{DampQchi}
|\delta\chi_{k_*}|\propto{\rm exp}(-Ht)\,.
\end{equation}
When $|\frac{\dot\chi\ddot\phi}{M_p^2H}\delta\phi_{k_*}|/|\frac{k^2}{a^2}\delta\chi_{k_*}|$ reaches $O(1)$,
$\delta\chi_{k_*}$ begins to grow due to the exponential growth of the source term, 
\begin{equation}\label{ExpQchi}
|\delta\chi_{k_*}|\propto|\delta\phi_{k_*}|\propto\exp\Big[\lambda_{k_*}H t\Big]\,,
\end{equation}
As long as the mode is within the resonant band, the exponential growth continues.
However, it cannot last indefinitely because $\phi$ stops evolving at $\phi\approx\phi_e$. 
The duration of the exponential growth of $\delta\phi_{k}$ is estimated in 
Eq.~(\ref{dt_phi}) in Appendix~\ref{ResonanceInExpand}.
For $k=k_*$, it gives
\begin{equation}
H\delta t_{\phi_{k_*}}\approx\ln\Big(\sqrt{\frac{4(1+Q)}{9P_0}}\Big)\,,
\end{equation}
where 
\begin{equation}
P_0=\Big(\frac{2f_aH}{\dot\phi_0}\Big)^2\,,\quad Q=2\frac{\Lambda^4}{\dot\phi_0^2}\,.
\end{equation}

\begin{figure}[h!]
\centering
\includegraphics[width=0.48\textwidth]{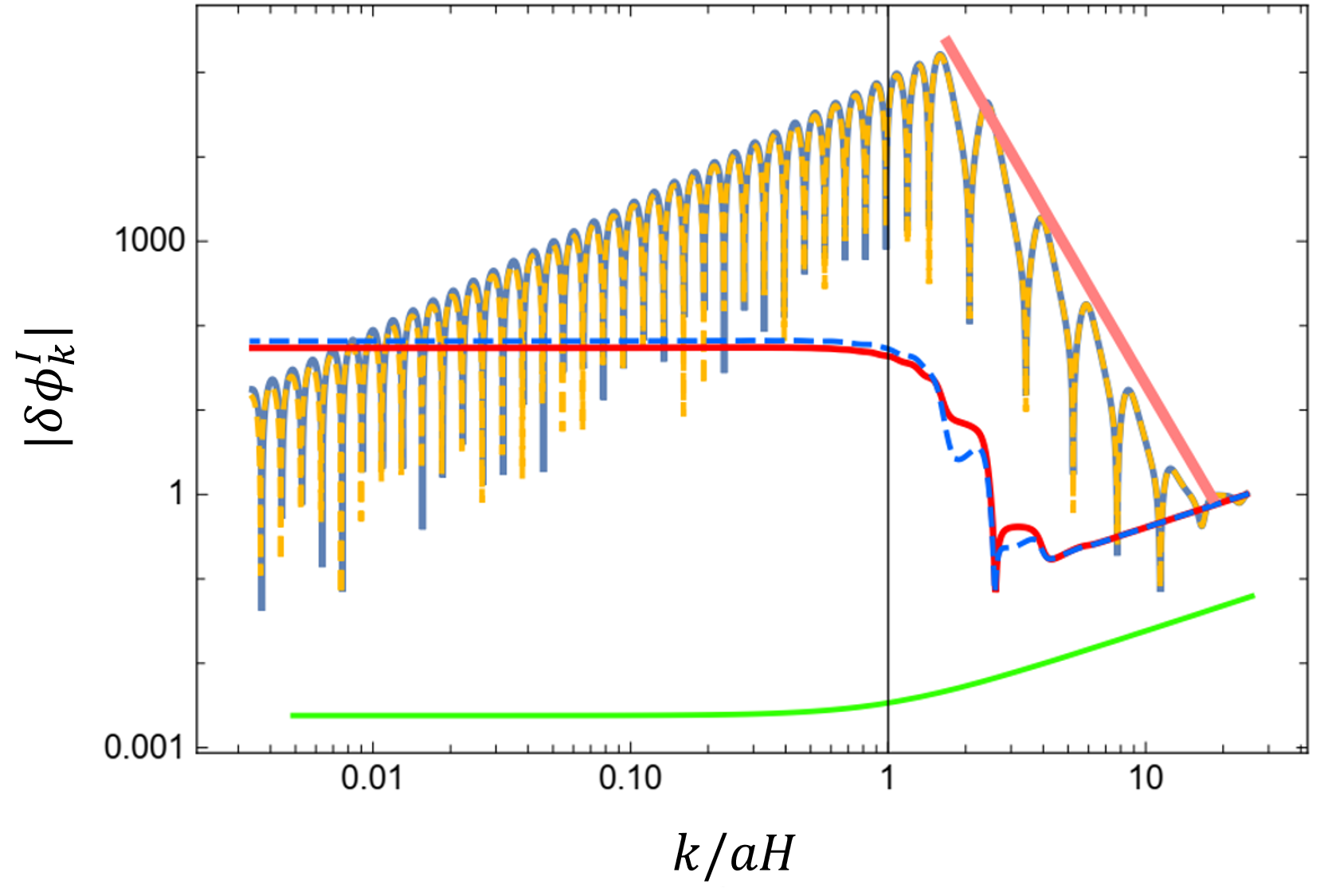}
\caption{Induced Resonance of $k_{*}$ mode with parameter choice: $g=-0.05$, $\xi=-0.1g$, $\alpha=0.55$, $\Lambda_0=4\times10^{-4}M_p$, $f_a=5\times10^{-3}M_p$, $V_0=3\times10^{-11}M_p^4$ and $k_{*}=10^{12}{\rm Mpc^{-1}}$.
The parameters are chosen to satisfy the observed perturbation amplitude $A_s\sim 10^{-9}$ as well as to realize the induced resonance.
The Hubble parameter during the resonance is about $H\simeq3.1\times10^{-6}M_p$. 
The blue and red solid lines are the numerical solutions of $\delta\phi_{k_*}$ and $\delta\chi_{k_*}$, respectively. 
The yellow and light blue dashed lines are their approximated solutions from Eq.~\eqref{EoMQphi} and \eqref{EoMQchi}. The pink line depicts the approximate analytical behavior given in Eq.~\eqref{ExpQphi}. 
The green line represents $\delta\chi_k$ for a mode $k$ that does not get the induced amplification.
 The vertical axis is the amplitude normalized with respect to the perturbation amplitude at the time when the $k_{*}$ mode enters the resonant band.}
\label{fig:FieldPert}
\end{figure}

By setting the mode functions to be in the Bunch-Davies vacuum initially,
\begin{equation}
\lim_{\tau\rightarrow-\infty} a\delta\phi_k=\lim_{\tau\rightarrow-\infty}a\delta\chi_k=\frac{e^{-ik\tau}}{\sqrt{2k}}\,,
\end{equation}
we numerically solve Eqs.~\eqref{BGKG} and \eqref{PertEoM}. The results for a particular set of model parameters
are shown in Fig.~\ref{fig:FieldPert}.
We see that $\delta\chi_k$ follows the usual dampened feature for modes $k$ that do not stay in the resonant band
and becomes frozen after Hubble exit.
As for the $k_{*}$ mode, $\delta\phi_{k_*}$ is amplified exponentially, reaches a maximum, and decays after $\phi$ stops rolling at $\phi_e$.
On the other hand, $\delta\chi_{k_*}$ evolves as given by Eqs.~\eqref{DampQchi} and \eqref{ExpQchi},
  showing the induced resonant amplification inside the Hubble horizon and frozen after Hubble exit. 

\subsection{Curvature perturbation spectrum}\label{PowerSpectrum}

Having solved the evolution of the field fluctuations up to Hubble exist and given the background dynamics, 
we can now compute the power spectrum of the primordial comoving curvature perturbation $P_{\calR}(k)$
use the $\delta N$ formalism \cite{sasaki1996general, sasaki1998super, lyth2005general, naruko2011conservation, abolhasani2019delta}.
The $\delta N$ formalism states that the final value of the conserved comoving curvature perturbation 
$\calR_c$ in the adiabatic limit on superhorizon scales is given by 
\begin{equation}\label{DeltaNDefine}
\calR_c({\bm x})=N({\bm x},t,t_f)-\bar{N}(t,t_f)=\delta N({\bm x},t)
\end{equation}
where $N$ is the number of $e$-folds from $t$ to $t_f$ locally determined from the homogeneous background evolution,
with $t_f$ being a cosmic time when the adiabatic limit is achieved and $t$ being any time before $t_f$ which is 
usually taken to be the time of horizon exit of the scale of interest.

 In our model we have 
\begin{align}\label{DeltaN}
\delta N({\bm x},t_i)=& \Big(\frac{\partial N}{\partial \phi}\delta\phi 
+\frac{\partial N}{\partial \chi}\delta\chi +\frac{1}{2}\frac{\partial^2 N}{\partial \phi^2}\delta\phi^2 \\
&+ \frac{\partial^2 N}{\partial \phi\partial\chi}\delta\phi\delta\chi 
+\frac{1}{2}\frac{\partial^2 N}{\partial \chi^2}\delta\chi^2\Big)_{t=t_i}\,, \nonumber
\end{align}
where $\delta\phi$ and $\delta\chi$ are the fluctuations evaluated on flat slicing at $t=t_i$.
Since the initial $\phi$-dominated stage is essentially the same as the standard single field inflation,
let us focus on the stage when and after $\phi$ stops evolving.
Then we may choose $t_i$ to be some time after the horizon exit when $\delta\phi$ has decayed out.
This implies we may ignore $\delta\phi$ at $t=t_i$ and hence
\begin{equation}
	\label{dN2nd}
\delta N({\bm x},t_i)= \Big(\frac{\partial N}{\partial \chi}\delta\chi +\frac{1}{2}\frac{\partial^2 N}{\partial \chi^2}\delta\chi^2
\Big)_{t=t_i}\,.
\end{equation}
In order to evaluate the above, we need to know the number of $e$-folds $N$ counted backward in time from 
the end of inflation to the time $t=t_i$. Then we find that the contribution of $\phi$ to $N$ is 
small, of $O((\phi-\phi_e^2)/M_p^2)$ at most, because $\phi$ becomes non-dynamical.
Thus we obtain
\begin{equation}
N = \frac{\chi ^2}{2 M_p^2} + \chi \frac{V_0 + g \Lambda _0^3 \phi_e}{\xi M_p^2 \Lambda_0^3} \\
+ O\big((\phi-\phi_e)^2/M_p^2\big)\,,
\label{N2nd}
\end{equation}
where we set $\chi=0$ to be the end of inflation for simplicity.

Using the above result, the power spectrum for $\calR$ is found as 
\begin{equation}\label{SpectrumFromDelChi}
P_{\calR}(k)=\frac{k^3}{2\pi^2}\Big|\frac{\partial N}{\partial\chi}\Big|^2|\delta\chi_k|^2(t_i)
\simeq\frac{H^2}{8\pi^2M_p^2\epsilon_{\chi\chi}}\mathcal{A}^2(k)\,,
\end{equation}
where an approximate expression for $\mathcal{A}^2(k)$ is given by 
\begin{align}
	\label{Aapproximate}
	\begin{split}
		\mathcal{A}^2(k)&=1+\mathcal{A}^2(k_*)\exp\Big(-\frac{\ln^2(k/k_*)}{2\Delta^2}\Big)\,,
	\end{split}
\end{align}
where $\Delta=\ln(k_+/k_-)/(2\sqrt2)$. A detailed derivation is given in Appendix.~\ref{ResonanceInExpand}. 
For the parameters given in the caption of Fig.~\ref{fig:FieldPert}, we find $\Delta=0.245$ and $\mathcal{A}^2(k_*)=10^{6.3}$.
Since we have the standard almost scale-invariant spectrum $P_{\calR}=A_s(k/k_p)^{n_s-1}$ outside the
resonant band, where $A_s$ is the amplitude of power spectrum and $n_s$ is the spectral index 
   at pivot scale $k_p\simeq0.05{\rm Mpc^{-1}}$ \cite{Akrami:2018odb},
   We may parametrize the spectrum over the whole range of $k$ as
$P_{\calR}=A_s(k/k_p)^{n_s-1}\mathcal{A}^2(k)$ with $\mathcal{A}^2(k)$ given by Eq.~(\ref{Aapproximate}).

Finally, let us evaluate the non-Gaussianity of the curvature perturbation. 
From Eq.~(\ref{N2nd}), the local non-Gaussianity is evaluated as
\begin{equation}\label{chi-nG}
\frac{3}{5}f_{NL}^{\rm local}=\dfrac{\partial_{\chi\chi}N}{2(\partial_\chi N)^2}=\epsilon_{\chi\chi}\,.
\end{equation}
Hence the local non-Gaussianity of the curvature perturbation turns out to be small. Nevertheless,
since $\delta\chi$ is proportional to $\delta\phi$ which is exponentially enhanced from its vacuum state,
a non-negligible intrinsic non-Gaussianity of $\delta\chi$ is expected to arise. 
We plan to come back to the issue of the intrinsic non-Gaussianity in the near future.

\begin{figure}[h!]
\centering
\includegraphics[width=0.45\textwidth]{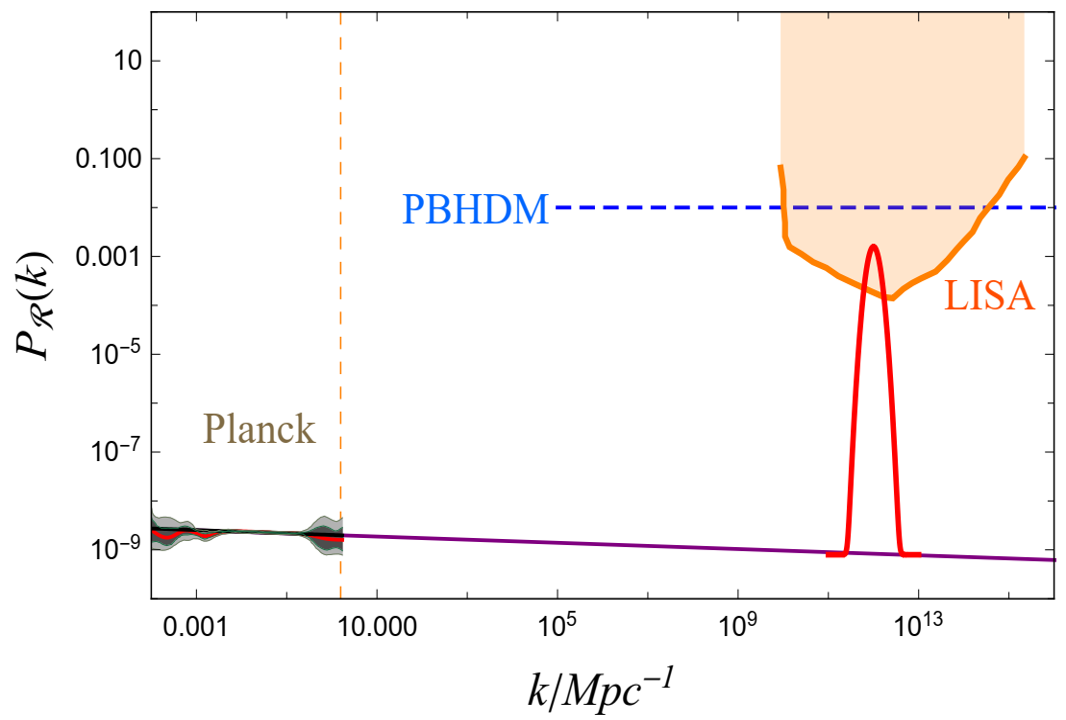}
\caption{The power spectrum of the comoving curvature perturbation with an enhancement 
	due to the induced resonance that fits the CMB constraint from the Planck data\cite{Akrami:2018odb}.
The model parameters are the same as those of Fig.~\ref{fig:FieldPert}.
The blue dashed line indicates the amplitude that would produce an appropriate amount of PBHs
that can account for cold dark matter \cite{Green:2020jor}. 
The orange curve is the expected sensitivity curve of LISA obtained for the induced GWs in our scenario\cite{Chluba:2019nxa}.}
\label{fig:PowerSpectrum}
\end{figure}

Before concluding this section, we emphasize that it is the induced $\delta\chi$ that determines the spectrum,
but $\delta\phi$ does not explicitly show up in the result.
Interestingly, on the other hand, as we discuss in Sec.~\ref{GWs}, the induced GWs are found to be
directly determined by $\delta\phi$.

\section{PBH formation}\label{PBH}

Now we study the abundance of PBHs due to the amplification in the primordial power spectrum on small scales.
 Assuming that the universe has reached the adiabatic limit when it is radiation-dominated,
 the comoving curvature perturbation is frozen on superhorizon scales. As the scale enters the Hubble horizon,
 some regions may have a large positive curvature that corresponds to a large positive 
 density perturbation on the comoving slice.  
 In the context of the local Friedmann equation with the curvature term $K/a^2$ being space-dependent,
 the relation between the curvature perturbation $\calR_c$ and  the density contrast $\delta=\delta\rho/\bar\rho$
  at linear order on superhorizon scales is given as
 \begin{align}
\frac{K({\bm x})}{a^2}=-\frac{2}{3}\frac{{}^{(3)}\Delta \calR_c}{a^2}=\frac{3}{2}H^2\delta\,.
\label{curvature-delta}
 \end{align}
 As the universe with a large positive curvature is equivalent to a closed universe where the expansion eventually stops and
the contraction starts, it is expected that a Hubble size region with a sufficiently large positive curvature will
collapse to a black hole. This is how a PBH forms.
It is then customarily to define a threshold $\delta_c$ in terms of $\delta$ extrapolated 
from the superhorizon expression (\ref{curvature-delta}) in linear theory to the horizon crossing $k=aH$.
Thus at horizon entry, we have $\delta=(4/9)\calR_c$ in Fourier space.
Although recently there have been substantial progress in studies of PBH formation criteria \cite{Nakama:2013ica},
here we stick to the simplest criterion and defer more thorough studies for future.

As an entire Hubble region becomes a PBH, it is also convenient to relate the comoving scale $k$
to the mass of a PBH which is equal to the mass of the corresponding Hubble region at the time of horizon-entry.
\begin{equation}
	\label{horizonmass}
M=\frac{4\pi}{3}\rho H^{-3}=\frac{1}{2GH}=\frac{4\pi M_p^2a}{k_M}\,.
\end{equation}
To estimate the abundance of PBHs, one usually assumes that the primordial curvature perturbation follows the Gaussian
statistics. In our model, we have seen that it is actually the case. 
Then the fraction of the energy density that turns into PBHs of mass $M$ at the time of formation is
estimated by integrating the probability distribution function for $\delta>\delta_c$,
\begin{equation}\label{MassFunc}
\beta(M)=\gamma{\rm Erfc}\Big[\frac{\delta_c}{\sqrt{2}\sigma_M}\Big]\,,
\end{equation}
where Erfc[$\cdots$] is the complementary error function and $\gamma\simeq0.2$ is a correction 
factor \cite{Carr:1975qj}, and $\sigma_{M}$ is the variance
of the density fluctuations at the PBH mass $M$ associated with scale $k_{M}$.
The variance $\sigma_{M}$ is estimated as 
\begin{equation}\label{MassDev}
\sigma_M^2=\int_{0}^{\infty}\frac{dk}{k}W^2(k/k_M)\frac{16}{81}(k/k_M)^4P_{\calR}(k)\,,
\end{equation}
where we have adopted the Gaussian window function $W(x)={\rm exp}(-x^2/2)$. 
For subtleties in the choice of window functions, see \cite{Tokeshi:2020tjq,Gow:2020bzo}.

It is well known that the observed almost scale-invariant part of the scalar curvature power spectrum is too small
to produce PBHs. Thus the spectrum must be enhanced on scales of interest, which is provided by
the resonant amplification in the current case. 
Given the function $\beta(M)$,  one can compute the fraction of PBHs against the total DM density at present.
It is given as\cite{sasaki2018primordial}
\begin{align}\label{FracDM}
f_{{\rm PBH}}(M) \equiv \frac{\Omega_{{\rm PBH}}}{\Omega_{DM}} = 1.52 &\times 10^8 \Big(\frac{\gamma}{0.2}\Big)^{1/2} \Big(\frac{g_{*,{\rm form}}}{106.75}\Big)^{-1/4} \nonumber \\ &\times \Big(\frac{M}{M_{\odot}}\Big)^{-1/2} \beta(M) ~,
\end{align} 
where $g_{*,{\rm form}}$ is the total relativistic degrees of freedom when PBHs form and $M_{\odot}=2\times10^{33}$g is the solar mass.
 The relation between the PBH mass $M$ and the corresponding comoving wavenumber $k_M$ may be expressed as
\begin{equation}\label{MassWaveNum}
M\simeq M_{\odot}\Big(\frac{\gamma}{0.2}\Big)\Big(\frac{g_{*,{\rm form}}}{106.75}\Big)^{-1/6}\Big(\frac{k_{M}}{1.6\times10^{6}{\rm Mpc^{-1}}}\Big)^{-2}\,.
\end{equation}

\begin{figure}[h!]
\centering
\includegraphics[width=0.47\textwidth]{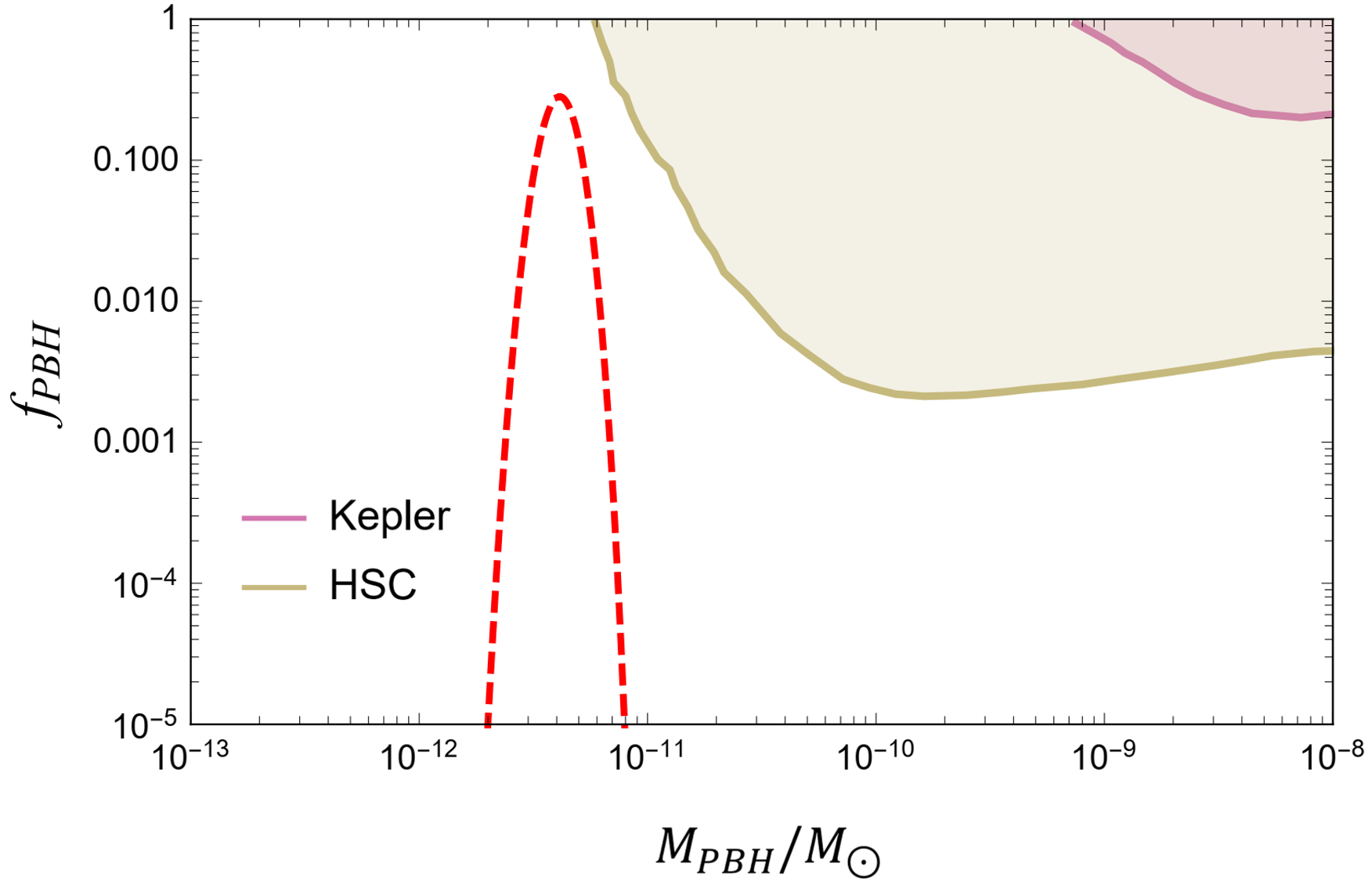}
\caption{The mass spectrum of PBH $f_{{\rm PBH}}$. The model parameters are the same as those of Fig.~\ref{fig:FieldPert}.
 The colored areas depict the constraints from two microlensing experiments, Kepler (pink) \cite{Griest:2013aaa} and 
HSC (light green)\cite{sugiyama2020wave}.}
\label{fig:PBH}
\end{figure}

For the model parameters of Fig.~\ref{fig:FieldPert}, the density fraction, $f_{{\rm PBH}}$, 
produced by the induced resonance as dipicted in Fig.~\ref{fig:PBH} is shown
for the parameter choice: $\gamma=0.2$, $g_{*,{\rm form}}\simeq100$ \cite{carr2010new} and
 $\delta_c=0.37$ \cite{sasaki2018primordial}. 
In this case, the mass spectrum is centered around $4\times10^{-12}M_{\odot}$ and PBHs constitute 
DM fraction of $f_{{\rm PBH}} \simeq 0.3$. 
This is within the current microlensing constraints, Kepler and HSC (Hyper Suprime-Cam) \cite{Carr:2020gox}.

Finally, we mention that since we are working in the perturbative regime, we have the basic theoretical
constraint $P_{\calR}\ll 1$.  It is of interest to see the consequences for a range of
model parameters that requires a non-perturbative treatment. We leave it for future studies. 

\section{Induced gravitational waves}\label{GWs}

In this section, we present a complete analysis of the induced GWs throughout the whole evolution of 
the very early universe. Details are deferred to Appendix~\ref{GWRev}. 
We first discuss GWs generated during the radiation-dominated era.

\subsection{Radiation-dominated era}

Choosing the Newton gauge for the scalar-type perturbations, the perturbed metric reads
\begin{equation}\label{NewtonGauge}
ds^2=-a^2(1+2\Psi)d\tau^2+a^2\Big[(1+2\Phi)\delta_{ij}+h_{ij}\Big]dx^idx^j ~,
\end{equation}
where $\Psi$ is the Newton potential, and $\Phi$ is the curvature perturbation. $h_{ij}$ the tensor (transverse-traceless) perturbation. In the absence of scalar-type anisotropic stress, we have $\Psi=-\Phi$.
Here we consider induced GWs, which originates from $h_{ij}$ to the second order sourced by the first-order scalar perturbations.

Decomposing $h_{ij}$ into Fourier modes, we have
\begin{equation}
	h_{ij}(\tau,{\bm x}) = \sum_{\lambda = +,\times} \int \frac{\mathrm{d}^{3} {\bm k}}{(2 \pi)^{3/2}}e^{i {\bm k} \cdot {\bm x}} h^\lambda_{{\bm k}}(\tau) e^\lambda_{ij}({\bm k})\,,
\end{equation}
where  $e^\lambda_{ij}$ is the orthonormal polarization tensor, and $\lambda=+,~\times$ represent two polarizations respectively.
The equation for the Fourier modes becomes
\begin{equation}\label{FouierGWeq}
	h^{\lambda''}_{{\bm k}}(\tau) + 2 \mathcal{H} h^{\lambda'}_{{\bm k}}(\tau) + k^2 h^\lambda_{{\bm k}}(\tau) 
	= S^\lambda_{{\bm k}}(\tau)\,,
\end{equation}
where the source term $S_{{\bm k}}^{\lambda}(\tau)$ is given by
\begin{align}\label{RadSource}
\begin{split}
S_{{\bm k}}^{\lambda}(\tau)&
=2\int\frac{d^3{\bm p}}{(2\pi)^{3/2}}\textbf{e}^{\lambda}({\bm k},{\bm p})\Big[2\Phi_{{\bm p}}(\tau)\Phi_{{\bm k}-{\bm p}}(\tau)+\\
&\Big(\Phi_{{\bm p}}(\tau)+\frac{\Phi'_{{\bm p}}(\tau)}{\mathcal{H}}\Big)
\Big(\Phi_{{\bm k}-{\bm p}}(\tau)+\frac{\Phi'_{{\bm k}-{\bm p}}(\tau)}{\mathcal{H}}\Big)\Big]\,,
\end{split}
\end{align}
where $\mathcal{H}=a'/a$ is the comoving Hubble parameter and 
$\textbf{e}^{\lambda}({\bm k},{\bm p})\equiv e_{lm}^{\lambda}({\bm k})p_lp_m$.
The solution may be formally expressed in terms of the retarded Green function, $g_{{\bm k}}(\tau,\tau')$,
\begin{equation}
	h^\lambda_{{\bm k}}(\tau)
	=\frac{1}{a(\tau)}\int^\tau_{-\infty} \mathrm{d} \tau_1 a(\tau_1) g_{\bm k}(\tau,\tau_1) S^\lambda_{{\bm k}}(\tau_1)\,.
	\label{Gfcnsol}
\end{equation}

  During the radiation-dominated era, $\Phi$ is given in terms of the conserved comoving curvature perturbation as
\begin{equation}\label{BardCur}
\Phi_{{\bm k}}(\tau) = \frac{2}{3} T(k\tau) \calR_{c,{\bm k}}\,,
\end{equation}
where $T(k\tau)$ is the transfer function, the explicit expression of which can be  found in the literature.
Using the retarded Green function $g_{{\bm k}}(\tau,\tau')=\frac{1}{k}{\rm sin}(k(\tau-\tau'))\Theta(\tau-\tau')$, 
in the radiation-dominated universe, the power spectrum of the induced GWs during radiation-dominated epoch 
is obtained as~\cite{ananda2007cosmological, baumann2007gravitational, garcia2017gravitational, bartolo2019testing}:
\begin{align}\label{RadPower}
P_h(k,\tau) =& \int_{0}^{+\infty} dy \int_{|1-y|}^{1+y} dx \Big[ \frac{4y^2 -(1+y^2-x^2)}{4xy} \Big]^2 \nonumber\\
& \times P_{\calR}(kx) P_{\calR}(ky) F(k\tau,x,y) ~,
\end{align}
where an explicit expression for $F(z,x,y)$ is given in Appendix.~\ref{GWRev}.
The resultant energy density spectrum $\Omega_{\rm GW}(f)h_0^2$ at present is 
shown by the orange line in Fig. ~\ref{fig:GW}. There the expected sensitivity curve of LISA is
also shown for comparison.

We can first use the analytical formula in \cite{pi2020gravitational} to 
estimate the maximum of the induced GW spectrum,
%\begin{align}
%&\Omega_\mathrm{GW}(k_*)h^2\nonumber\\
%&\quad\approx1.6\times10^{-5}A_s^2\left(\frac{k_*}{k_p}\right)^{2(n_s-1)}
%\left(0.41\sqrt{2\pi}\Delta\mathcal{A}^2(k_*)\right)^2,
%\nonumber\\
%&\quad\approx5.5\times10^{-11},
%\end{align}
\begin{align}\nonumber
&\,\Omega_\mathrm{GW,0}h^2\\\nonumber
\approx&\,1.6\times10^{-5}\times2.6\left(A_s^2\left(\frac{k_*}{k_p}\right)^{2(n_s-1)}\sqrt{2\pi}\Delta\mathcal{A}(k_*)^2\right)^2\\
\approx&\,2.9\times10^{-11}.
\end{align}
where we have adopted the values $\Delta=0.245$, $\mathcal{A}^2(k_*)=10^{6.3}$, $k_*=10^{12}~\text{Mpc}^{-1}$, $k_p=0.05~\text{Mpc}^{-1}$, $A_s^2=2.2\times10^{-9}$, and $n_s=0.96$, which are listed in Section~\ref{PowerSpectrum}.
%As the value of the width $\Delta=0.245$ indicates, the induced resonant mechanism leads to a fairly broad scale amplification of the curvature perturbation spectrum, $\delta k\sim 1.5k_*$. 

 Here, we should also mention that, the GWs generated during this epoch reflects the behavior of $\delta\chi$ that accounts for
 the curvature perturbation from inflation, while the role of $\delta\phi$ which is the origin of the resonant amplification,
 is hidden from view  in the sense that its behavior cannot be reconstructed from the final curvature perturbation spectrum.
 On the other hand, as we shall see in the next subsection, 
 the GWs induced during inflation directly reflects the behavior  of $\delta\phi$.

\subsection{Inflationary era}

Here we consider GWs produced inside the Hubble horizon during inflation.
As the field fluctuations in the resonance band are exponentially enhanced, they can generate a non-negligible amount of GWs \cite{Cai:2019jah}.
On sub-Hubble scale during inflation, a convenient choice of gauge is flat slicing.
The metric and relevant equations are summarized in Appendix~\ref{MInflation}.

The source term $S_{{\bm k}}^{\lambda}(\tau)$ for the tensor perturbation is given by  
\begin{align}\label{InfSource}
S_{{\bm k}}^{\lambda}(\tau) =&
 \frac{2}{M_p^2} \int \frac{d^3{\bm p}}{(2\pi)^{3}} \textbf{e}^{\lambda}({\bm k},{\bm p}) \delta\phi_{{\bm p}}(\tau) \delta\phi_{{\bm k}-{\bm p}}(\tau) \nonumber\\ 
&+(\phi\leftrightarrow\chi)\,.
\end{align}
Since we are interested in the generation of GWs on sub-Hubble scales, the background may be approximated by a de Sitter (dS) space,
with $a(\tau)=-1/(H\tau)$ and $\mathcal{H}=a'/a=-1/\tau$ ($-\infty<\tau<0$). 
The retarded Green function in dS space is given by \cite{biagetti2013enhancing}
\begin{align}
g_{{\bm k}}(\tau,\tau') =& \frac{1}{k^3\tau'^2} \Big[ -k(\tau-\tau') ~ {\rm cos} k(\tau-\tau') \nonumber\\
& + (1+k^2\tau\tau'){\rm sin} k(\tau-\tau')\Big] \Theta(\tau-\tau')\,.
\label{dSGreen}
\end{align}

\begin{figure}[h!]
\centering
\includegraphics[width=0.48\textwidth]{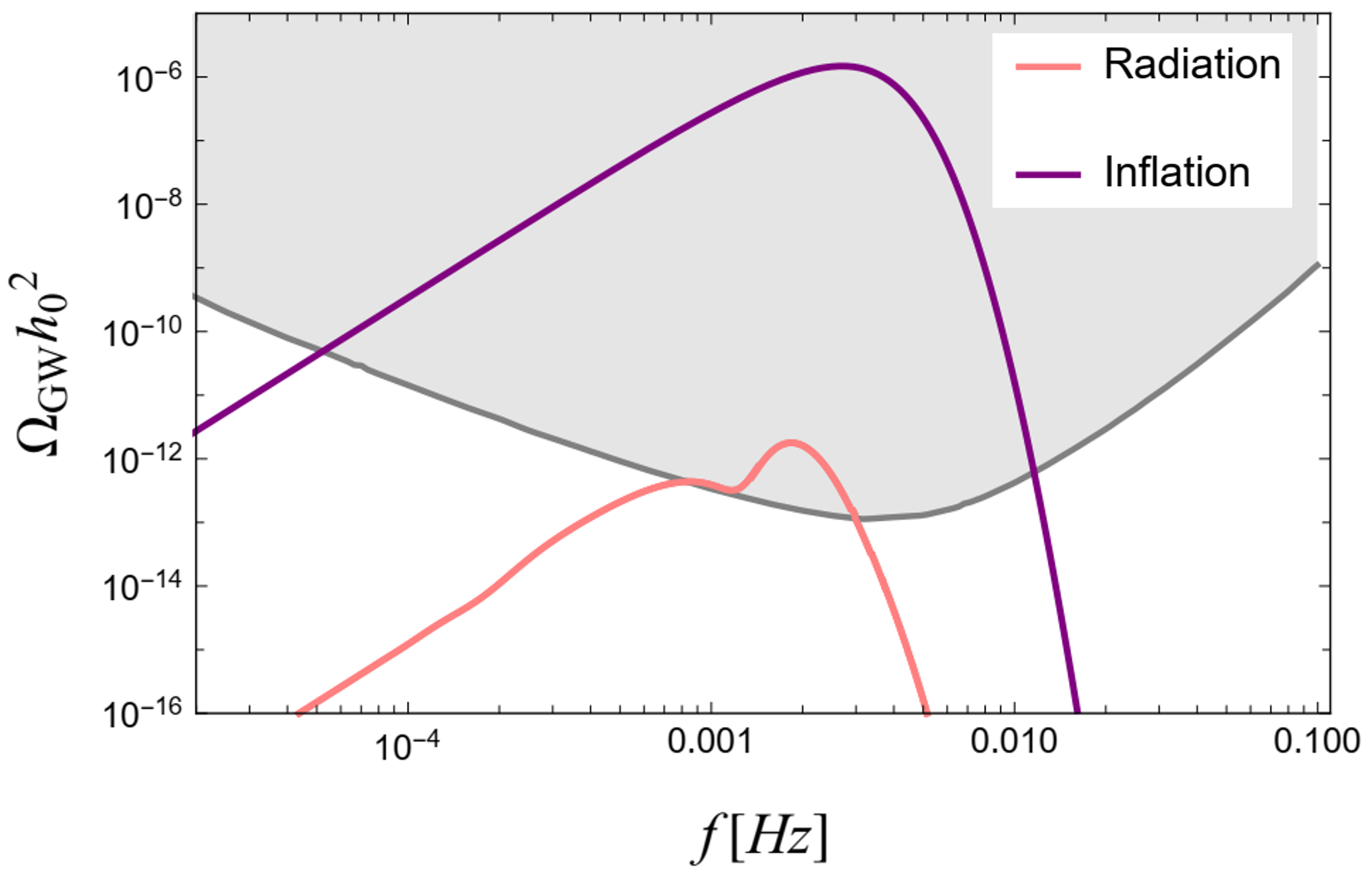}
\caption{The energy spectra of GWs produced in our model.
The pink line denotes the induced GWs from the radiation-dominated era,
and the purple line denotes the induced GWs from inflationary era.
The gray region is within the expected sensitivity of LISA \cite{Audley:2017drz}. 
The model parameters are the same as those given in the caption of Fig.~\ref{fig:FieldPert}. }
\label{fig:GW}
\end{figure}

Using the above Green function, one can immediately obtain the solution as given in Eq.~(\ref{Gfcnsol}).
Then the power power spectrum can be readily expressed as \cite{biagetti2013enhancing}
\begin{align}\label{IFGWPower}
P_h(k,\tau_{\rm end})&=\frac{4}{\pi^4M_p^4}k^3\int_{0}^{\infty}dp p^6\int_{-1}^{1}d{\rm cos}\theta\space{\rm sin}^4\theta \nonumber\\
&\times\Big|\int_{\tau_0}^{\tau_{end}}d\tau_1g_k(\tau_{end},\tau_1)(\delta\phi_p(\tau_1)\delta\phi_{|{\bm k}-{\bm p}|}(\tau_1) \nonumber\\
&+\delta\chi_{p}(\tau_1)\delta\chi_{|{\bm k}-{\bm p}|}(\tau_1))\Big|^2 ~,
\end{align}
where $\tau_{end}\approx0$ represents the conformal time at the end of inflation. 
It may be useful to note that the ultraviolet behavior of $P_h(k)$ is given by
\begin{equation}\label{IFGWUltraviolet}
P_h(k,\tau_{\rm end})\propto P_{\phi}(k),\quad \text { for } k\gg k_*.
\end{equation}
We have numerically integrated the above, and after taking into account the thermal history of the universe,
we have obtained the energy density spectrum $\Omega_{\rm GW}(f)h_0^2$ of the induced GWs.
The result is the purple line shown in Fig.~\ref{fig:GW}. As easily seen, the contribution from the
induced GWs during inflation is far larger than that from the induced GWs during radiation-dominated era.

It is important to note that, the induced GWs from inflation is dominated by $\delta\phi$,
since it is much larger $\delta\chi$ on sub-Hubble scales as shown in Fig.~\ref{fig:FieldPert}. 
In the conventional case of inflation where there is no resonant effect, the induced GWs from inflation is suppressed by the square of 
the slow-roll parameter compared to those generated during radiation-dominated era, since
$P^2_\phi(k)/M_p^4\sim \epsilon^2 P^2_\calR(k)$, where $\phi$ is the inflaton and $\epsilon=-\dot H/H$ is the slow-roll parameter.
However, when the resonance occurs, $\delta\phi$ which is responsible for the resonance, is much larger than $\delta\chi$
which is induced by the enhanced $\delta\phi$. This can result in the dominance of the inflationary GWs.
In fact, we can estimate the ratio of the amplitude for the GW spectrum induced from inflationary era and radiation-dominated era as
\begin{align}
	\frac{\Omega_{\rm GW}^{\rm Inf}}{\Omega_{\rm GW}^{\rm Rad}}
	&\sim 
	\epsilon_{\chi\chi}^2\frac{P^2_\phi(k_*)}{P^2_\chi(k_*)}
	\nonumber\\\label{Omratio}
	%&\sim 
	%\epsilon_{\chi\chi}^2\exp\Big(4(\lambda_{k_*}+1)H(\delta t_{\phi_{k_*}}-\delta t_{\chi_{k_*}})\Big)\\
	%\nonumber\\
	&\sim \epsilon_{\chi\chi}^2\Big(\frac{\mu_{k_*}^2}{Q}\frac{3M_p^2H^2}{2f_a\xi\Lambda_0^3}\Big)^4.
\end{align}
For the model parameters given in the caption of Fig.~\ref{fig:FieldPert},
${\Omega_{\rm GW}^{\rm Inf}}$ is a factor of $O(10^5)$ larger than ${\Omega_{\rm GW}^{\rm Rad}}$.

The above result is encouraging. With future observations of cosmological GW backgrounds, we may be able to deduce the information of $\delta\phi$ during inflation. Together with observations of PBHs, we may be able to disclose a new physics behind inflation.

\section{Conclusion}\label{Con}

In this paper, we put forward a novel mechanism that can realize a controllable instability of an entropy perturbation during inflation 
which is eventually converted into the curvature perturbation. 
This can be achieved by a two-field model with one field possessing an axion-monodromy like oscillating potential and the other 
with a hilltop-like potential. By solving the perturbation equations stage by stage, 
we found that the oscillatory dynamics of the axion-like field $\phi$ can lead to parametric resonance of its field fluctuations. 

Although the growth of these perturbations does not necessarily give rise to enhanced curvature perturbations as these modes 
are amplified only on sub-Hubble scales, the fluctuations induced in the second field $\chi$ could also be dramatically enhanced
 to produce large curvature perturbations.
 In fact, in our model where the inflationary trajectory makes a turn from the $\phi$-direction to the $\chi$-direction after
 $\phi$ is stuck at one of the oscillatory minima, the induced fluctuations in $\chi$ which are initially entropic are converted into
 adiabatic perturbations after the turn.
 As a result, the curvature perturbation displays a manifest resonant growth on small scales, up to an amplitude
 enough to produce PBHs abundantly.
 
  Another intriguing result is that the amplified fluctuations in $\phi$, which has no direct effect on the curvature perturbation,
  can produce tensor perturbations during inflation when they are on sub-Hubble scales.
These produced tensor perturbations will eventually form a cosmological GW background, and their amplitude can be
much larger than those induced by the curvature perturbation during the radiation-dominated universe.
As an example, we have explicitly demonstrated the case in which the induced GWs from inflation dominated over
those from the radiation-dominated era, with the spectrum large enough to be observed by future GW detectors like LISA.

Given the aforementioned phenomenological success, we would like to comment on the implications of the proposed
 model that could inspire certain follow-up studies. From a theoretical perspective, it seems it is important to understand better the physics of resonance in the early universe, that is, to understand how the whole system reacts when one of the physical degrees of freedom experiences a resonance
and what the condition for effective resonances induced in other degrees of freedom is.
As in our model, such induced resonances may be more important in producing phenomena that can be used to probe the theory. 

Another issue is non-Gaussianity. We have not studied non-Gaussianties in our model in detail. But as we
mentioned in section, we expect a large non-Gaussianity in the scalar field perturbations on sub-Hubble scales
because of their amplifications as well as of their highly non-linear self-interactions.
As it is well known, the production of PBHs as well as the induced GWs may be significantly affected by a non-Gaussianity in the
distribution function of the curvature perturbation~\cite{Cai:2018dig}. This issue definitely deserves detailed studies.

Our mechanism can greatly enhance the induced GWs, which leaves a better chance of detecting 
such small-scale primordial perturbations by stochastic GWs instead of the PBHs. %This is opposite to the model studied in \cite{Cai:2018dig}, which suppresses the induced GW spectrum by non-Gaussian distribution of the scalar perturbation when fixing the PBH abundance. 
This is because the resonance greatly enhances the induced GWs for a given abundance of PBHs, which implies that for a possible detection of the stochastic gravitational waves, for instance 
the recent NANOGrav 12.5 year result~\cite{Arzoumanian:2020vkk}, there might be no significant amount of PBHs 
that contributions to a substantial portion of either the cold dark matter~\cite{DeLuca:2020agl} or 
supermassive black holes~\cite{Vaskonen:2020lbd}.

Finally, we mention that the results derived from our model has successfully demonstrated 
the importance of joint analyses on PBHs and GWs.  Along with the combined observational data, 
which may be able to reveal a surprising new physics behind inflation that wouldn't be visible in large scale observations like CMB. 

%\section*{Acknowledgments}
{\it Acknowledgments.--}
We are grateful to Chao Chen, Jinn-Ouk Gong, Jerome Quintin, Yi Wang, Dong-Gang Wang, Ziwei Wang and Hao Yang for stimulating discussions. SP thanks the hospitality of School of Physics and Electronics, Hunan Normal University during his visit. YFC is supported in part by the NSFC (Nos. 11722327, 11653002, 11961131007, 11421303), by the CAST-YESS (2016QNRC001), by the National Youth Talents Program of China, and by the Fundamental Research Funds for Central Universities. MS is supported in part by JSPS KAKENHI Nos. 19H01895 and 20H04727. SP is supported in part by JSPS Grant-in-Aid for Early-Career Scientist No. 20K14461. Both SP and MS are supported by the World Premier International Research Center Initiative (WPI Initiative), MEXT, Japan. 
All numerics are operated on the computer clusters {\it LINDA} \& {\it JUDY} in the particle cosmology group at USTC.

\appendix

\section{Basic formula for multi-field inflation} \label{MInflation}

In this appendix, we review the basic formulas for multi-field inflation \cite{seery2005primordial}. 
We consider a general multi-component scalar field $\phi^{I}$ ($I=1,2,\cdots,n$),
\begin{equation}
\begin{split}
S=&\frac{1}{2} \int \mathrm{d}^{4} x \sqrt{-g}\Bigl[M_p^2R\\
&\quad-\mathcal{G}_{I J}(\phi)g^{\mu\nu} \partial_{\mu} \phi^{I} \partial_{\nu} \phi^{J}-2 V(\phi)\Bigr]\,,
\end{split}
\end{equation}
where $R$ is the Ricci scalar and $\mathcal{G}_{I J}$ is the field space metric. 
For the simplest case of a canonical scalar field  $\mathcal{G}_{I J}=\delta_{I J}$.
Performing the $(3+1)$ decomposition of the metric,
\begin{equation}
ds^2=-N^2dt^2+\gamma_{ij}(dx^i+N^idt)(dx^j+N^jdt)\,.
\end{equation}
the action reduces to
\begin{align}
\begin{split}
&S=-\frac{1}{2} \int d^4x N \sqrt{\gamma}\left(\mathcal{G}_{I J} \gamma^{i j} \partial_{i} \phi^{I} \partial_{j} \phi^{J}-2 V(\phi^{I})\right)\\
&+\frac{M_p^2}{2} \int d^4x N \sqrt{\gamma}\left(K_{i j} K^{i j}-K^{2}+\frac{1}{M_p^2N^2}\mathcal{G}_{I J} v^{I} v^{J}\right),
\end{split}
\end{align}
where $K_{ij}$ is the extrinsic curvature of $t=const.$ slices, and $v^{I}=\dot{\phi}^{I}-N^{j} \partial_{j} \phi^{I}$.

As well-known, the lapse function $N$ and the shift vector $N^i$ serve as Lagrange multipliers.
Taking the variations with respect to $N$ and $N^i$, we get the Hamiltonian and momentum constraints, respectively,
\begin{align}
&\mathcal{G}_{I J} \gamma^{i j} \partial_{i} \phi^{I} \partial_{j} \phi^{J}+2 V
+\frac{1}{N^{2}}\mathcal{G}_{I J} v^{I} v^{J}
\nonumber\\
&\qquad\qquad+M_p^2\left(K_{i j} K^{i j}-K^{2}\right)=0\,.
\label{consteqs}\\
&D_j\left[K_{i}^{j}-K \delta_{i}^{j}\right]=\frac{1}{M_p^2N} \mathcal{G}_{I J} v^{I} \partial_{i} \phi^{J}\,,
\nonumber
\end{align}
where $D_i$ is the covariant derivative with respect to the spatial metric $\gamma_{ij}$. 

We now consider linear perturbation in a spatially flat cosmological background, 
\begin{equation}\label{bgmetric}
N=1\,,\quad N^i=0\,, \quad \gamma_{ij}=a^2(t)\delta_{ij}\,,\quad \phi^I=\phi^I(t)\,.
\end{equation}
At linear order, we set $N=1+A$, $N_i=\partial_i B$, and $\phi^I=\phi^I(t)+\delta\phi^I$,
while keeping the spatial metric to be flat, $\gamma_{ij}=a^2\delta_{ij}$. Namely, we
choose the spatially flat gauge. With this choice of gauge, 
the constraint equations (\ref{consteqs}) can be solved for $A$ and $B$ as
\begin{align}\label{Constraints}
\begin{split}
A&=\frac{1}{2 HM_p^2} \mathcal{G}_{I J} \dot{\phi}^{I} \delta\phi^{J}\,,\\
B&=\partial^{-2}\psi\,;\\
&\psi=\frac{a^{2}}{2 HM_p^2} \mathcal{G}_{I J}\left(\delta\phi^{I} \ddot{\phi}^{J}
-\dot{\phi}^{I} \dot{\delta\phi}^{J}-\frac{\dot{H}}{H} \dot{\phi}^{I} \delta\phi^{J}\right)\,,
\end{split}
\end{align}
where $\partial^{-2}$ is an integral operator such that $B=\partial^{-2}\psi$ satisfies $\partial^2B=\psi$.

After expanding the action to second order in perturbation and plugging Eq.~\eqref{Constraints} into it,
with the help of the background equations, we find the action for $\delta\phi^I$,
\begin{align}\label{S2}
\begin{split}
S_{2}&=\frac{1}{2} \int \mathrm{d} t \mathrm{d}^{3} x a^{3}\Big(\mathcal{G}_{I J} \dot{\delta\phi}^{I} \dot{\delta\phi}^{J}\\
&-\frac{1}{a^{2}} \mathcal{G}_{I J} \partial_i \delta\phi^{I} \partial^i \delta\phi^{J}-\mathcal{M}_{I J}^2 \delta\phi^{I} \delta\phi^{J}\Big)\,,
\end{split}
\end{align} 
where the effective mass square matrix is given by 
\begin{equation}
\mathcal{M}_{I J}^2=V_{, I J}-\frac{1}{M_p^2a^{3}} \frac{\mathrm{d}}{\mathrm{d} t}\left(\frac{a^{3}}{H} \dot{\phi}_{I} \dot{\phi}_{J}\right)\,.
\end{equation}
Taking the variation of Eq.~(\ref{S2}), we obtain
\begin{align}
\begin{split}
\ddot{\delta\phi}^{I}&+3 H \dot{\delta\phi}^{I}+\frac{k^2}{a^{2}} \delta\phi^{I}+ \\
&\sum_{J}\left[V_{, I J}-\frac{1}{M_{p}^{2} a^{3}} \frac{d}{d t}\left(\frac{a^{3}}{H} \dot{\phi}_{I} \dot{\phi}_{J}\right)\right] \delta\phi^{J}=0\,,
\end{split}
\end{align}
where we have assumed $\mathcal{G}_{I J}=\delta_{IJ}$ and performed the Fourier transformation.
Applying the above to our model, we obtain the perturbation equations of motion as given in Eq.~\eqref{PertEoM}. 

\section{Parametric resonance in our model}\label{ResonanceInExpand}
\begin{figure}[h!]
	\centering
	\includegraphics[width=0.47\textwidth]{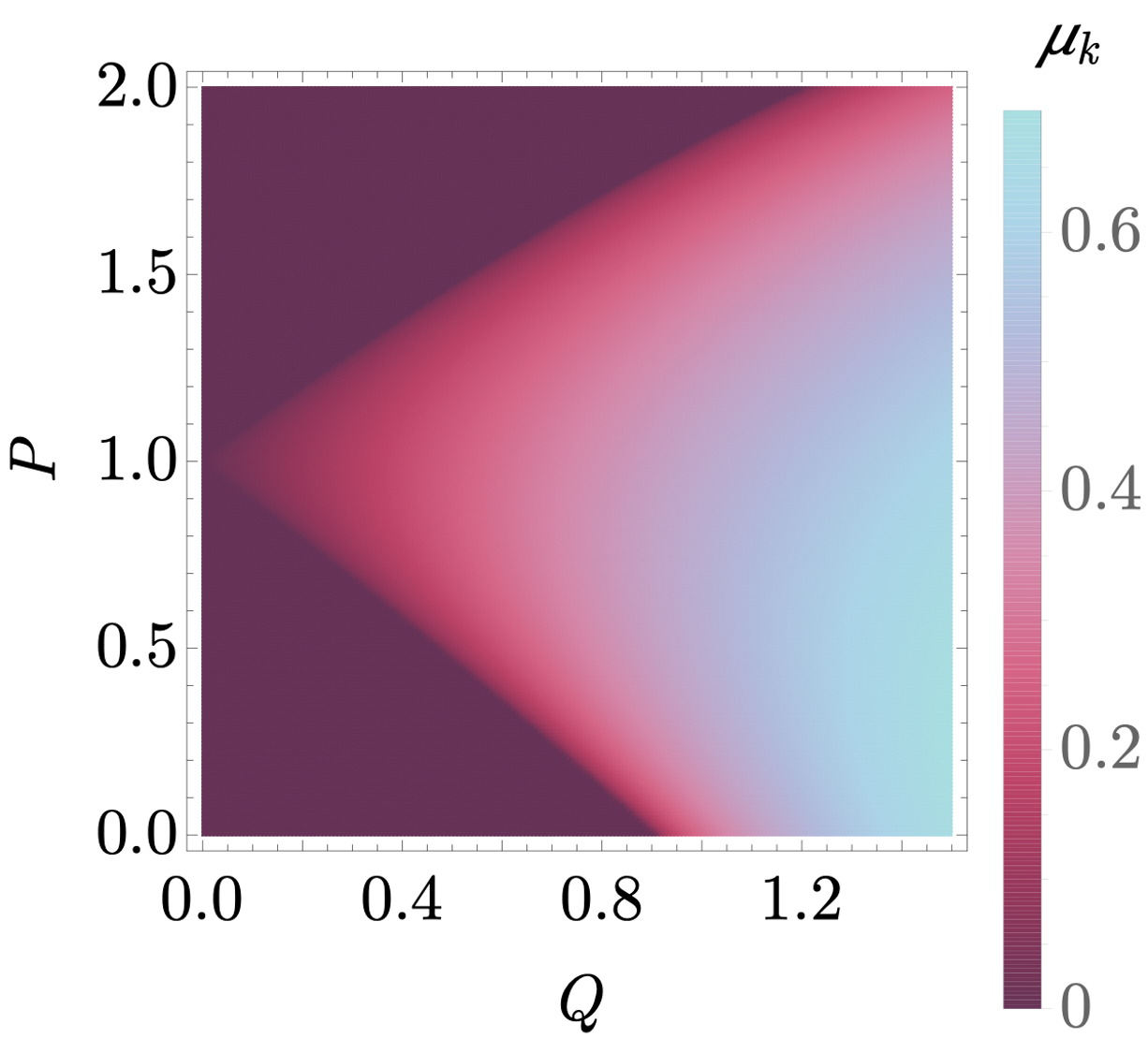}
	\caption{The stability/instability chart for the standard Mathieu equation, where $\mu_k$ is the Floquet number. $P$ and $Q$ are the equation parameters.}
	\label{fig:MathieuBand}
\end{figure}

In this appendix, we give a detailed analysis for parametric resonance in our model which is captured in Eq.~\eqref{EoMDPhi}. 
After introducing the new variable $dz=\dot\phi_0dt/(2f_a)$ and setting $\delta\phi_k=a^{-3/2}\Phi_k$,
we get an approximate equation in the form of the Mathieu equation,
\begin{equation}
\frac{d^2\delta\Phi_k}{dz^2}+[P-2Q\cos(2z)]\delta\Phi_k=0\,,
\end{equation}
where
\begin{align}
\begin{split}
P=P_0\Big(\frac{k^2}{a^2H^2}-\frac{9}{4}\Big)\,&,~~P_0=\Big(\frac{2f_aH}{\dot\phi_0}\Big)^2\,,\\
Q=Q_0\Big(\frac{\Lambda}{\Lambda_0}\Big)^4\,&,~~Q_0=2\frac{\Lambda_0^4}{\dot\phi_0^2}\,.
\end{split}
\end{align}
In our case, $P$ and $Q$ are not constants. Instead, they will evolve along with the new variable $z$. 
But, in the current analysis, we just consider the case where the maximum of $Q$ is of $O(1)$.
In the instability region of our interest, we have $\delta\Phi_k(z+\pi)=e^{\pi(\mu+i)}\delta\Phi_k(z)$, where $\mu$ is the Floquet number.
The stability/instability chart is presented in Fig.~\ref{fig:MathieuBand} with the values of $\mu$ in gradient colors.

From Fig.~\ref{fig:MathieuBand}, we see that the instability band at $Q\lesssim O(1)$ is $1-Q\lesssim P\lesssim 1+Q$. It proves useful to introduce a reference mode $k_*$ which crosses the horizon just at time $t_e$. Thus, for $k<k_*$, the duration of $\delta\phi_k$ resonance is 
\begin{equation}
H\delta t_{\phi_k}=\ln\Big(\frac{2}{3}\sqrt{\frac{1+Q}{P_0}}\Big)\,.
\end{equation}
For $k>k_*$, the duration of $\delta\phi_k$ resonance is 
\begin{equation}\label{dt_phi}
H\delta t_{\phi_k}=\ln\Big(\frac{k_*}{k}\Big)+\ln\Big(\sqrt{\frac{4(1+Q)}{9P_0}}\Big)\,.
\end{equation}
During the resonance, the behavior of $\delta\phi_k$ is
\begin{equation}
|\delta\phi_k|\propto a^{-3/2}\exp(\mu_kz)=\exp\Big[\lambda_k Ht\Big]\,,
\end{equation} 
where 
\begin{equation}
	\lambda_k=\mu_{k}\frac{|g|\Lambda_0^3}{6H^2f_a}-\frac{3}{2}\,.
	\label{growthrateB}
\end{equation}
Then, defining the amplification factor for $\delta\phi_k$ as
\begin{equation}\label{Ampdef}
\delta\phi_k(t_{k,e})=\tilde{\mathcal{A}}_\phi(k)\delta\phi_k(t_{k,b})\,,
\end{equation}
where $t_{k,b}$ and $t_{k,e}$ are the beginning and ending epochs of the resonance, respectively,
we obtain
\begin{equation}
\tilde{\mathcal{A}}^2_\phi(k)=\exp\Big[2\lambda_kH\delta t_{\phi_k}\Big]\,.
\label{Ampphi}
\end{equation}

The exponential amplification of $\delta\phi_k$ will induce a similar amplification of $\delta\chi_k$ 
which can be directly related to the coming curvature perturbation according to the $\delta N$ formalism.
 In section~\ref{PertAmplification}, we notice that 
 when $|\frac{\dot\chi\ddot\phi}{M_{p}^2H}\delta\phi_{k_*}|/|\frac{k^2}{a^2}\delta\chi_{k_*}|\gtrsim O(1)$,
 $\delta\chi_k$ will show the same exponential amplification behavior as $\delta\phi_k$ and the induced resonance will occur. 
 Then setting $\delta\chi_k\propto\delta\phi_k\propto\exp[\lambda_k Ht]$,  
 Eq.~\eqref{EoMQchi} gives the ratio $|\delta\chi_k|/|\delta\phi_k|$ at the end of resonance as
 \begin{equation}\label{chiphiratio}
 F\equiv \frac{|\delta\chi_k|}{|\delta\phi_k|}\simeq\frac{1}{(\lambda_k+3/2)^2}\frac{\dot\chi\ddot\phi}{M_p^2H}
 	=\frac{Q}{\mu_{k}^2}\frac{2\xi\Lambda_0^3f_a}{3M_p^2H^2}\,,
 \end{equation}
where we set $k^2/a^2=9/4$ as the end of resonance time.
From eq.~(\ref{Ampdef}) and using the fact that $|\delta\chi_k|\simeq|\delta\phi_k|$ at the beginning of the resonance,
we obtain
\begin{align}
\begin{split}
|\delta\chi_k(t_{k,e})|
&= F|\delta\phi_k(t_{k,e})|
= F\tilde{\mathcal{A}}_\phi |\delta\chi_k(t_{k,b})|
\\
&\simeq F\tilde{\mathcal{A}}_\phi \exp[H\delta t_{\phi_k}] |\delta\chi_k^{\rm vac}(t_{k,e})|\,,
\end{split}
\end{align}
where $\delta\chi_k^{\rm vac}(t_{k,e})$ is the vacuum amplitude at the end of the resonance if there were no resonance.
Thus, the amplification factor for $\delta\chi_k$ relative to its vacuum value is found as
\begin{equation}
\tilde{\mathcal{A}}_\chi^2(k)\simeq F^2\exp[2(\lambda_k+1)H\delta t_{\phi_k}]\,.
\label{Ampchi_k}
\end{equation}

Given the amplification factor for $\delta\chi_k$, we now consider the $\delta\chi_k$ spectrum.
 In order to capture the behavior of the power spectrum, we approximate it by a log-normal function near the peak,
 \begin{equation}\label{lognorm}
 P(k)\propto \exp\left[-\frac{\ln^2(k/k_*)}{2\Delta^2}\right]\,.
 \end{equation}
 Since we know the value of the amplification factor, the remaining task is to compute the width $\Delta$,
 which can be expressed as
\begin{equation}
\Delta=\frac{1}{2\sqrt2}\ln\Big(\frac{k_+}{k_-}\Big)\,,
\label{Deltadef}
\end{equation} 
where $k_{\pm}$ are the wavenumbers such that they satisfy
$\tilde{\mathcal{A}}_\chi^2(k_-)=\tilde{\mathcal{A}}_\chi^2(k_+)=\tilde{\mathcal{A}}_\chi^2(k_*)/e$.

For modes $k>k_*$, we easily find
\begin{equation}
\ln\Big(\frac{k_+}{k_*}\Big)\simeq\frac{1}{2(\lambda_{k_+}+1)}\,.
\label{kplus}
\end{equation}
For modes $k<k_*$, it is a bit complicated task. We first note that the duration of the resonance is approximately independent of $k$, but
$\mu_k$ is smaller for smaller $k$ because $Q$ is an increasing function of time and we expect $\mu_k\propto Q$ for $Q\lesssim 1$ from 
Fig.~\ref{fig:MathieuBand}. As an epoch at which the resonance is most effective is given by $P\sim1$, which means $k\sim aH P_0$,
the slow time variation of $Q$ can be translated to the slow variation of the amplification with respect to $k$,
\begin{equation}\label{tdepmu}
\begin{split}
\frac{d}{d\ln k}\ln\mu_k
&=\frac{d}{Hdt}\ln\mu_k\approx\frac{d}{Hdt}\ln Q
\\
&=\frac{4\alpha\dot\phi}{HM_p}\frac{\Lambda}{\Lambda_0}
\end{split}
\end{equation}
This implies
\begin{equation}
\mu(k)=\mu(k_*)(\frac{k}{k_*})^{\beta}\,;\quad
\beta=4\alpha\frac{\Lambda_0}{\Lambda}\frac{\dot\phi}{HM_p}\,.
\end{equation}
Applying the condition $\tilde{\mathcal{A}}_\chi^2(k_-)=\tilde{\mathcal{A}}_\chi^2(k_*)/e$ to Eq.~(\ref{Ampphi}),
and using the above result, we obtain
\begin{equation}
\ln\Big(\frac{k_*}{k_-}\Big)=\frac{1}{\beta}\ln\Big(\frac{\ln B}{\ln B-1}\Big)\,,
\label{kminus}
\end{equation}
where 
\begin{equation}
B=\exp\Big[H\delta t_{\chi_{k_*}}\mu_{k_*}\frac{\dot\phi_0}{Hf_a}\Big]\,.
\end{equation}

\begin{figure}[h!]
	\centering
\includegraphics[width=0.47\textwidth]{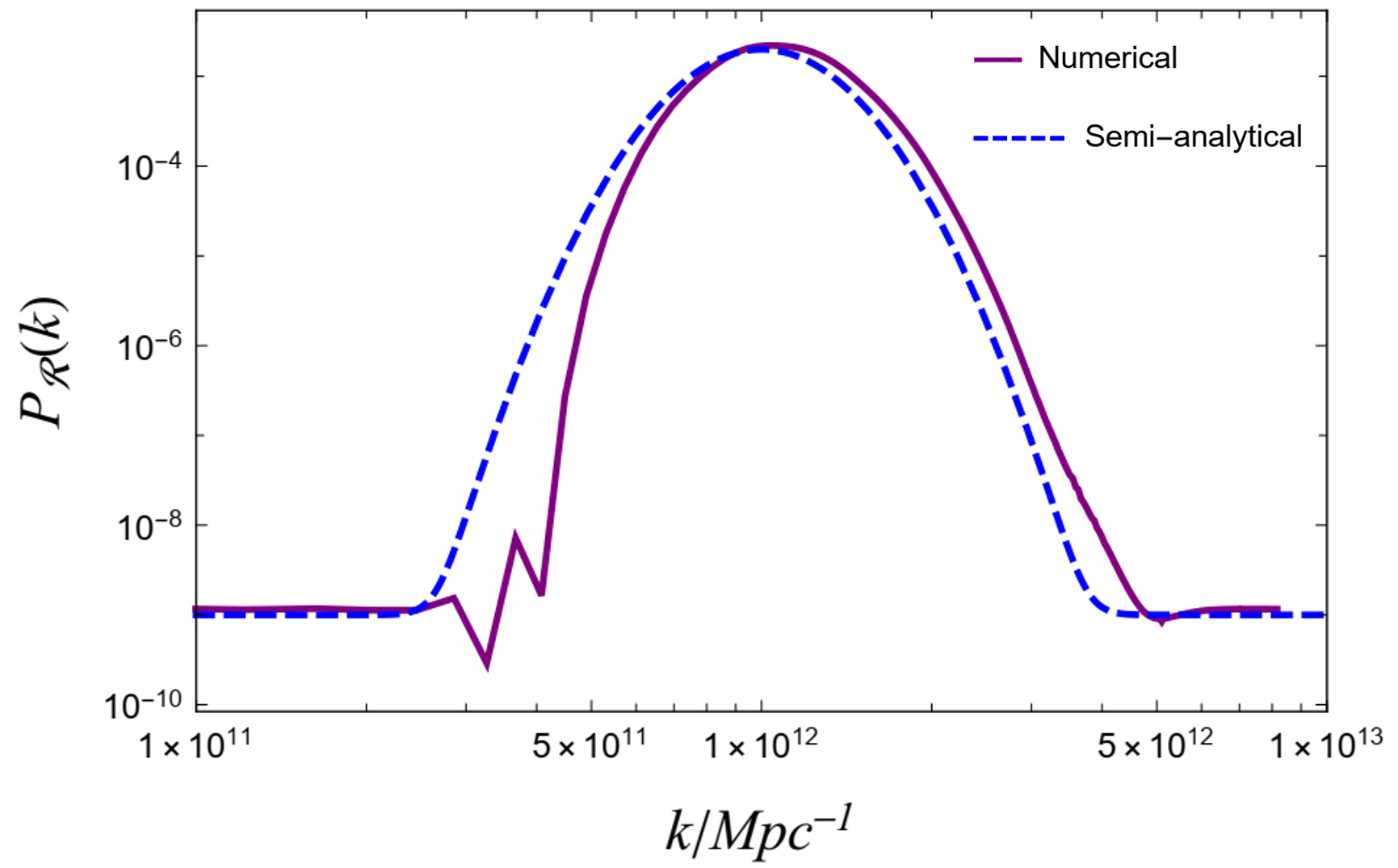}
	\caption{The power spectrum calculated via two methods. The result from numerical calculation is plotted in the purple solid line. The blue dashed line shows the semi-analytical result from Eq.~\eqref{SemiSpectrum}. In the given example, 
		numerically $\Delta=0.245$, semi-analytically $\Delta=0.276$ and $\tilde{\mathcal{A}}^2(k_*)=10^{6.3}$.}
	\label{fig:PowerSpectrum2}
\end{figure}

Thus Eq.~(\ref{Deltadef}) together with Eqs.~(\ref{kplus}) and (\ref{kminus}) determines the width $\Delta$.
The resulting near-peak spectrum for $\delta\chi_k$, and hence that for the comoving curvature 
perturbation is, with a fairly good accuracy, proportional to
\begin{equation}\label{SemiSpectrum}
\mathcal{A}^2(k)=1+\tilde{\mathcal{A}}_{\chi}^2(k_*)\exp\Big(-\frac{\ln^2(k/k_*)}{2\Delta^2}\Big)\,.
\end{equation}
For the model parameters used in the text, the numerically computed power spectrum and the above approximate spectrum
are compared in Fig.~\ref{fig:PowerSpectrum2}. As we see from it, the approximate formula fits the numerical result reasonablly well.

\section{Basic formula for induced GWs}\label{GWRev}

In this appendix, we review the basic formulas for computing the induced GWs \cite{ananda2007cosmological,baumann2007gravitational,boyle2008probing}.
 In the cosmological background, the equation of motion for the transverse-traceless (TT) tensor perturbation $h_{ij}(\tau,{\bm x})$ 
 ($\partial_{i}h^{ij}=h^i{}_{i}=0$) with a source term $\mathcal{S}_{ij}(\tau,{\bm x})$ is 
\begin{equation}\label{hEoM}
h_{ij}''(\tau,{\bm x})+2\mathcal{H}h_{ij}'(\tau,{\bm x})-\nabla^2h_{ij}(\tau,{\bm x})
=-4\hat{T}^{\ell m}_{ij}\mathcal{S}_{\ell m}(\tau,{\bm x})\,,
\end{equation} 
where $\mathcal{H}=a'/a$ is the comoving Hubble parameter. 
The operator $\hat{T}^{\ell m}_{ij}$ is a TT projection tensor,
\begin{equation}
\hat{T}^{\ell m}_{ij}({\bm x})
=\sum_{\lambda=+,\times}\int\frac{d^3{\bm k}}{(2\pi)^{3/2}}e^{i{\bm k}\cdot{\bm x}}e_{ij}^{\lambda}\big({\bm k})e^{\lambda,\ell m}({\bm k})\,,
\end{equation}
where $\lambda=+,\times$ denote the two polarizations of GWs. 
The polarization tensor $e_{ij}^{\lambda}({\bm k})$ can be expressed
 in terms of the unit vectors orthogonal to ${\bm k}$,  $e^{(a)}_i({\bm k})$ ($a=1,2$) as 
\begin{align}
\begin{split}
e_{ij}^{+}({\bm k})=\frac{1}{\sqrt{2}}[e^{(1)}_{i}({\bm k})e^{(1)}_{j}({\bm k})-e^{(2)}_i({\bm k})e^{(2)}_j({\bm k})]\,,\\
e_{ij}^{\times}({\bm k})=\frac{1}{\sqrt{2}}[e^{(1)}_{i}({\bm k}) e^{(2)}_{j}({\bm k})+e^{(2)}_i({\bm k})e^{(1)}_j({\bm k})]\,.
\end{split}
\end{align}
Then the equation of motion in Fourier space becomes
\begin{equation}\label{FouierEoM}
\left[\frac{d^2}{d\tau^2} + 2 \mathcal{H}\frac{d}{d\tau}+ k^2 \right]h^\lambda_{{\bm k}}(\tau)
 = S^\lambda_{{\bm k}}(\tau)\,,
\end{equation}
where $h^\lambda_{\bm k}$ is defined by
\begin{equation}
h^\lambda_{{\bm k}}=e^{\lambda,\ell m}({\bm k})
\int \frac{d^3{\bm x}}{(2\pi)^{3/2}}e^{-i{\bm k}\cdot{\bm x}}h_{\ell m}(\tau,{\bm x})\,,
\end{equation}
and $S_{{\bm k}}^\lambda(\tau)$ by
\begin{equation}
S_{{\bm k}}^\lambda(\tau)=-4e^{\lambda, \ell m}({\bm k})\int\frac{d^3{\bm x}}{(2\pi)^{3/2}}e^{-i{\bm k}\cdot{\bm x}}S_{\ell m}(\tau,{\bm x})\,.
\end{equation}
The solution to Eq.~\eqref{FouierEoM} may be expressed in terms of the Green function,
\begin{equation}
\left[\frac{d^2}{d\tau^2} + 2 \mathcal{H}\frac{d}{d\tau}+ k^2 \right]g_{{\bm k}}(\tau,\tau_1) = \delta(\tau - \tau_1)\,.
\end{equation}
With an appropriate (usually the retarded) boundary condition specified to $g_{{\bm k}}$, the solution is given by
\begin{equation}
h^\lambda_{{\bm k}}(\tau)
=\int^\infty_{-\infty} \mathrm{d} \tau_1 g_{\bm k}(\tau,\tau_1) S^\lambda_{{\bm k}}(\tau_1)\,.
\end{equation}

Assuming the source is a spatially homogeneous, parity conserving random process,
the 2-point function of $h^\lambda_{\bm k}$ can be expressed as
\begin{equation}
\langle {h}^\lambda_{{\bm k}}(\tau){h}^s_{{\bm k}'}(\tau) \rangle 
= (2 \pi)^3 \delta^{\lambda s} \delta^{(3)}({\bm k} + {\bm k}') \frac{2 \pi^2}{2k^3} P_h(k, \tau)\,,
\end{equation}
where $P_h(k,\tau)$ is the dimensionless power spectrum of the tensor perturbation per unit logarithmic interval of $k$,
and we have used the fact that the power spectrum is the same for both polarizations,
\begin{equation}\label{key}
P_h(k)=\sum_{\lambda=+,\times} P^\lambda_h(k)=2P^\lambda_h(k)\,.
\end{equation}

In the case of GWs induced by the scalar perturbation during the radiation-dominated era,
the power spectrum at time $\tau$ is given by
\begin{align}\label{RadGW}
P^{\rm Rad}_h(k,\tau) =& \int_{0}^{+\infty} dy \int_{|1-y|}^{1+y} dx \Big[ \frac{4y^2 -(1+y^2-x^2)}{4xy} \Big]^2 \nonumber\\
& \times P_{\calR}(kx) P_{\calR}(ky) F(k\tau,x,y)\,,
\end{align}
where $P_{\calR}(k)$ is the spectrum of the conserved comoving curvature perturbation,
the function $F(z,x,y)$ is given by
\begin{align}
\begin{split}
F(z,x,y)&=\frac{4}{81}\frac{1}{z^2}[{\rm cos}^2(z)\mathcal{I}_c^2(x,y)\\
&+{\rm sin}^2(z)\mathcal{I}_s^2(x,y)+{\rm sin}(2z)\mathcal{I}_s(x,y)\mathcal{I}_c(x,y)] ~,
\end{split}
\end{align}
and the functions $\mathcal{I}_c$ and $\mathcal{I}_s$ by
\begin{align}
\begin{split}
\mathcal{I}_c(x,y)&=4\int_{1}^{\infty}dz_1(-z_1{\rm sin}z_1)\Big[2T(xz_1)T(yz_1)\\
&+[T(xz_1)+xz_1T'(xz_1)][T(yz_1)+yz_1T'(yz_1)]\Big]\\
\mathcal{I}_s(x,y)&=4\int_{1}^{\infty}dz_1(z_1{\rm cos}z_1)\Big[2T(xz_1)T(yz_1)\\
&+[T(xz_1)+xz_1T'(xz_1)][T(yz_1)+yz_1T'(yz_1)]\Big] \,,
\end{split}
\end{align}
with $T(z)$ being the transfer function,
\begin{equation}
T(z)=\frac{9}{z^2}\Big[\frac{{\rm sin}(z/\sqrt{3})}{z/\sqrt{3}}-{\rm cos}(z/\sqrt{3})\Big]\,.
\end{equation}

A stochastic background of the GWs can be characterized by its energy density fraction $\Omega_{\rm GW}$,
\begin{equation}
\Omega_{\text{GW}}(\tau,k) = \frac{1}{\rho_c(\tau)} \frac{\mathrm{d} \rho_{\text{GW}}(\tau,k)}{\mathrm{d} \ln k}
\,.
\end{equation}
where $\rho_c(\tau)=3M_p^2H^2(\tau)$ is the critical energy density at the conformal time $\tau$,
and $\rho_{\rm GW}(k)$ is the effective energy density spectrum of the GWs,
\begin{equation}
\rho_{\rm GW}(k)=\frac{M_p^2}{4}\frac{k^2}{a^2}P_h(k,\tau)\,.
\end{equation}

The induced GWs form a stochastic GW background today, and it is characterized by 
 $\Omega_{\rm GW}(\tau_0,k)$.  Taking account of the thermal history of the universe,
 the current energy density fraction of the induced GWs from the radiation-dominated era
 is given by \cite{Saito:2009jt}
\begin{equation}
\Omega^{\rm Rad}_{\rm GW}(\tau_0, k)=\Omega_{r,0}\Omega^{\rm Rad}_{\rm GW}(\tau_{\rm eq},k)\,,
\end{equation}
where $\tau_{\rm eq}$ is the time of radiation-matter equality and $\Omega_{r,0}$ is the present radiation energy density fraction.

As for the induced GWs from the inflationary era, 
the energy density spectrum at frequencies  is approximately given by 
%\dg{I didn't have time to check the numerical coefficient. Shi, can you do that?}

\begin{equation}
\Omega_ {\rm GW}^ {\rm Inf} (\tau_0, k)=\frac{ 1 }{ 12 }g_ {\rm eff } ^ {- 1 / 3 } \Omega _ { r , 0 } P_h( \tau _ {\rm end} , k)\,,
\end{equation}
where the modes are assumed to have entered the horizon during the radiation-dominated stage, $\tau_{\rm end}$ is the time at the end of inflation, and
$g_{\rm eff}$ is the effective number of degrees of freedom contributing to the radiation at that epoch.

\section{Vacuum stability}\label{Stable}

In order to demonstrate the validity of our model, it is very important for us to check the vacuum stability at $\phi=\phi_e$,
where the evolution in the $\phi$-direction stops.
Otherwise, there might occur phase transitions or multi-vacua quantum tunnelings as first touched in \cite{Zhou:2020stj}. 
In the present scenario, we need such probability to be negligibly small so that our perturbative treatment can be justified.
To be specific, a meta-stable vacuum may be regarded as stable if its lifetime is much longer than the age of our universe.
Thus let us evaluate the quantum tunneling probability of $\phi$ to decay to a lower minimum.

The vacuum decay rate per volume is estimated by
\begin{equation}\label{DecayRate}
\Gamma = \Gamma_0e^{-S_4[\phi]}\,,
\end{equation}
where $S_4$ is the Euclidean bounce action that connects the meta-stable vacuum to another vacuum with lower energy.
 $ \Gamma_0$ is related to the energy scale associated with the vacuum decay. Here for simplicity we assume
 $\Gamma_0\simeq H^4$, where $H$ is the Hubble parameter during inflation. We will see that the result is completely
 independent of this choice.
 
 For the lifetime of a vacuum to be longer than the age of the universe, we require
\begin{equation}\label{NowHubble}
\Gamma \ll H_0^4\,,
\end{equation}
where $H_0$ is the Hubble constant today. This requirement then reduces to the condition $S_4 > 520$.

In order to calculate the bounce action, we assume that the relevant part of the potential has remained the same until today,
\begin{equation}\label{PhiPoten}
\begin{split}
V(\phi) &= g\Lambda_0^3\phi + \Lambda^4(\phi_e)\cos(\frac{\phi}{f_a})
\\
&=g\Lambda_0^3\left(\phi-b_*(\phi_e)f_a\cos(\frac{\phi}{f_a})\right)
\,,
\end{split}
\end{equation}
and ignore the effect of gravity, which is valid at energy scales much smaller than the Planck scale.
We then Wick rotate to Euclidean time, $t\to i\tau$, and introduce the radial coordinate $r=\sqrt{{\bm x}^2+{\tau}^2}$.
Assuming the $O(4)$-symmetry, $\phi=\phi(r)$, the Euclidean equation of motion is
\begin{equation}\label{BounceEq}
\frac{d^2\phi}{dr^2} + \frac{3}{r}\frac{d\phi}{dr} 
-g\Lambda_0^3\left(1+b_*(\phi_e)\sin(\frac{\phi}{f_a})\right)= 0\,.
\end{equation}
We solve this equation with the boundary condition,
\begin{equation}\label{BoundCon}
\lim_{r \to \infty}\phi(r) = \phi_e\,,  \qquad  \frac{d\phi}{dr}\bigg|_{r=0} = 0\,.
\end{equation}
The bounce action of the $O(4)$-symmetric solution takes the form,
\begin{equation}
S_4[\phi]=2 \pi^2 \int_{0}^{\infty} dr\, r^3\bigg[\frac{1}{2}\bigg(\frac{d\phi}{dr}\bigg)^2+V(\phi)\bigg]\,.
\end{equation}

In the current situation, $b_*$ at $\phi=\phi_e$ is assumed to be larger than unity, $b_*(\phi_e)>1$, 
as mentioned in Sec.~\ref{PhysicalModel}, which corresponds to the large barrier limit. 
This allows us to simplify the bounce action to be
\begin{equation}
S_4 = \frac{6912}{\pi}\frac{f_{a}^3}{|g|\Lambda_0^3}\,.
\end{equation}
For the parameters we used in the text, we have $b_*(\phi_e)\simeq4.6$, and the above approximation gives  $S_4\simeq9\times10^7$. 
This is in good consistent with the numerical value $S_4=8.8\times10^7$. 
Now, as we see, the bounce action $S_4$ is exponentially larger than the estimated lower bound 520.
Thus the vacuum at $\phi=\phi_e$ is absolutely stable.

\bibliography{IRPBH20201028}

\end{document}